# Gravitational waves: Classification, Methods of detection, Sensitivities, and Sources

Kazuaki Kuroda,[†] Wei-Tou Ni[*] and Wei-Ping Pan[*]

[†]*Institute for Cosmic Ray Research, The University of Tokyo,*
*5-1-5, Kashiwanoha, Kashiwa, Chiba 277-8582, Japan*
kuroda@icrr.u-tokyo.ac.jp
[*]*Center for Gravitation and Cosmology (CGC), Department of Physics,*
*National Tsing Hua University, Hsinchu, Taiwan, 300, ROC*
weitou@gmail.com, d9722518@oz.nthu.edu.tw

After giving a brief introduction and presenting a complete classification of gravitational waves (GWs) according to their frequencies, we review and summarize the detection methods, the sensitivities, and the sources. We notice that real-time detections are possible above 300 pHz. Below 300 pHz, the detections are possible on GW imprints or indirectly. We are on the verge of detection. The progress in this field will be promising and thriving. We will see improvement of a few orders to several orders of magnitude in the GW detection sensitivities over all frequency bands in the next hundred years.

## 1. Introduction and Classification

Soon after the proposal of general relativity (GR), Einstein predicted the existence of gravitational waves (GWs) and estimated its strength from the wave equation he obtained in his 1916 paper on "Approximative Integration of the Field Equations of Gravitation".[1] Toward the end of his paper, he obtained the expression of the radiation $A$ of the system per unit time in general relativity as (equation (23) in his paper) $A = (\kappa/24\pi)$ $\Sigma_{\alpha\beta}(\partial^3 J_{\alpha\beta}/\partial\tau^3)^2$ with $J_{\alpha\beta}$ defined as the time-variable components of moment of inertia of the radiating system ($\kappa = 8\pi G_N$ in terms of Newtonian gravitational constant $G_N$).[a] He then continued that "This expression (for the radiation $A$) would get an additional factor $1/c^4$ if we would measure time in seconds and energy in Erg (erg). Considering $\kappa = 1.87 \cdot 10^{-27}$ (in units of cm and gm), it is obvious that $A$ has, in all imaginable cases, a practically vanishing value." Indeed at that time, possible expected source strengths and the detection capability had a huge gap. However, with the great strides in the advances of astronomy and astrophysics and in the development of technology, this gap is largely bridged. White dwarf was discovered in 1910 with its density soon estimated. Now we understand that GWs from white dwarf binaries in our Galaxy form a stochastic GW background ("confusion limit")[3] for space (low frequency) GW detection in GR. The first artificial satellite Sputnik was launched in 1957. However, at present the space GW missions are only expected to be launched in about 19 years later (~2034).[4]

---

[a]This radiation formula is corrected with the trace contribution of the moment of inertia subtracted and the overall factor replaced by $\kappa/80\pi$ [a factor 2 off compared with (37)] in Einstein's next paper on GWs.[2] With his correction, Einstein noticed that "This result shows that a mechanical system which permanently retains spherical symmetry cannot radiate…."

The existence of GWs is the direct consequence of GR and unavoidable consequence of all relativistic gravity theories with finite velocity of propagation. Maxwell's electromagnetic theory predicted electromagnetic waves. Einstein's general relativity and other relativistic gravity theories predict the existence of GWs. GWs propagate in space-time forming ripples of space-time geometry.

The role of GW in gravity physics is like the role of electromagnetic wave in electromagnetic physics. The importance of GW detection is twofold: (i) as probes to explore fundamental physics and cosmology, especially black hole physics and early cosmology, and (ii) as a tool in astronomy and astrophysics to study compact objects and to count them, complement to electromagnetic astronomy and cosmic ray (including neutrino) astronomy.

The existence of gravitational radiation is demonstrated by binary pulsar orbit evolution.[5,6] In GR, a binary star system would emit energy in the form of GWs. The loss of energy results in the shrinkage of the orbit and shortening of orbital period. Based on more than thirty-two years (from 1974 through 2006) of timing observations of the relativistic binary pulsar B1913+16, the cumulative shift of peri-astron time is over 43 s. The calculated orbital decay rate in general relativity using parameters determined from pulsar timing observations agreed with the observed decay rates. From this and a relative acceleration correction due to solar system and pulsar system motion, Weisberg, Nice and Taylor[6] concluded that the measured orbital decay to the GR predicted value from the emission of gravitational radiation is $0.997 \pm 0.002$ providing conclusive evidence for the existence of gravitational radiation as their previous papers. Kramer *et al.*[7] did an orbit analysis of the double pulsar system PSR J0737-3039A/B from 2.5 years of pulse timing observations and found that the orbit period shortening rate 1.252(17) agreed with the GR prediction of 1.24787(13) to 1.003(14) fraction. Freire et al.[8] analyzed about 10 years of timing data of the binary pulsar J1738+0333 and obtained the intrinsic orbital decay rate to be $(-25.9 \pm 3.2) \times 10^{-15}$, agreed well with the calculated GR value $(-27.7^{+1.5}_{-1.9}) \times 10^{-15}$ using the determined orbital parameters. Further precision and many more systems are expected in the future for observable GW radiation reaction imprint on the orbital motion.

The usual way of detection of gravitational wave (GW) is by measuring the strain $\Delta l/l$ induced by it. Hence gravitational wave detectors are usually amplitude sensors, not energy sensors. The detection of GWs can be resolved into characteristic frequencies. The conventional classification of gravitational-wave frequency bands, as given by Thorne[9] in 1995, was into (i) High-frequency band (1 Hz-10 kHz); (ii) Low-frequency band (100 µHz - 1 Hz); (iii) Very-low-frequency band (1 nHz -100 nHz); (iv) Extremely-low-frequency band (1 aHz - 1 fHz). This classification was mainly according to frequency ranges of corresponding types of detectors/detection methods: (i) Ground GW detectors; (ii) Space GW detectors; (iii) Pulsar timing methods; (iv) Cosmic Microwave Background (CMB) methods. In 1997, we followed Ref. [9] and extended the band ranges to give the following classification:[10,11]

     (i) High-frequency band (1-10 kHz);
     (ii) Low-frequency band (100 nHz - 1 Hz);
    (iii) Very-low-frequency band (300 pHz-100 nHz);
    (iv) Extremely-low-frequency band (1 aHz - 10 fHz).

Subsequently, we added the very-high-frequency band and the middle-frequency band for there were enhanced interests and activities in these bands. Recently we added the missing band (10 fHz – 300 pHz) and the 2 bands beyond to give a complete frequency classification of GWs as compiled in Table 1.[12-17]

Table 1. Frequency Classification of Gravitational Waves.[16-17]

| Frequency band | Detection method |
|---|---|
| Ultrahigh frequency band: above 1 THz | Terahertz resonators, optical resonators, and magnetic conversion detectors |
| Very high frequency band: 100 kHz – 1 THz | Microwave resonator/wave guide detectors, laser interferometers and Gaussian beam detectors |
| High frequency band (audio band)*: 10 Hz – 100 kHz | Low-temperature resonators and ground-based laser-interferometric detectors |
| Middle frequency band: 0.1 Hz – 10 Hz | Space laser-interferometric detectors of arm length 1,000 km – 60,000 km |
| Low frequency band (milli-Hz band)†: 100 nHz – 0.1 Hz | Space laser-interferometric detectors of arm length longer than 60,000 km |
| Very low frequency band (nano-Hz band): 300 pHz – 100 nHz | Pulsar timing arrays (PTAs) |
| Ultralow frequency band: 10 fHz – 300 pHz | Astrometry of quasar proper motions |
| Extremely low (Hubble) frequency band (cosmological band): 1 aHz – 10 fHz | Cosmic microwave background experiments |
| Beyond Hubble-frequency band: below 1 aHz | Through the verifications of inflationary/primordial cosmological models |

*The range of audio band normally goes only to 10 kHz.
†The range of milli-Hz band is 0.1 mHz to 100 mHz.

In Sec. 2, we give a brief introduction to GWs in general relativity. In Sec. 3, we review various methods of detection together with their typical/aimed sensitivities. In Sec. 4 we review various astrophysical and cosmological sources. In Sec. 5, we present an outlook.

## 2. Gravitational Waves (GWs) in General Relativity (GR)

The equations of motion of GR, i.e. the Einstein equation is

$$G_{\mu\nu} = \kappa \, T_{\mu\nu}, \tag{1}$$

where $T_{\mu\nu}$ is the stress-energy tensor and $\kappa = 8\pi G_N$. (We use the MTW[18] conventions with signature −2; this is also the convention used in [19]; Greek indices run from 0 to 3; Latin indices run from 1 to 3; the cosmological constant is negligible for treating the methods of GW detection and for evaluating GW sources at the aimed accuracy of this paper except in the Hubble frequency band and beyond the Hubble frequency band, and will be neglected in this treatment except in association with cosmological models.). Contracting the equations of motion (1), we have

$$R = - \, 8\pi \, G_N \, T, \tag{2}$$

where $T \equiv T_\mu{}^\mu$. Substituting (2) into (1), we obtain the following equivalent equations of motion

$$R_{\mu\nu} = 8\pi G_N \, [T_{\mu\nu} - (1/2)(g_{\mu\nu}T)]. \tag{3}$$

For weak field in the quasi-Minkowskian coordinates, we express the metric $g_{\alpha\beta}$ as

$$g_{\alpha\beta} = \eta_{\alpha\beta} + h_{\alpha\beta}, \quad h_{\alpha\beta} << 1. \tag{4}$$

Since $h_{\alpha\beta}$ is a small quantity, we expand everything in $h_{\alpha\beta}$ and linearize the results to obtain the linear approximation. For linearized quantities, we use the Minkowski metric $\eta_{\alpha\beta}$ to raise and lower indices without affecting the linearized results. The Riemann curvature tensor can be expressed as

$$R_{\alpha\beta\gamma\delta} = (1/2)(g_{\alpha\delta,\beta\gamma} + g_{\beta\gamma,\alpha\delta} - g_{\alpha\gamma,\beta\delta} - g_{\beta\delta,\alpha\gamma}) + g_{\mu\nu}(\Gamma^\mu{}_{\beta\gamma}\Gamma^\nu{}_{\alpha\delta} - \Gamma^\mu{}_{\beta\delta}\Gamma^\nu{}_{\alpha\gamma}). \tag{5}$$

With linearization, we have

$$R_{\alpha\beta\gamma\delta} = (1/2)(h_{\alpha\delta,\beta\gamma} + h_{\beta\gamma,\alpha\delta} - h_{\alpha\gamma,\beta\delta} - h_{\beta\delta,\alpha\gamma}) + \mathrm{O}(h^2), \tag{6}$$

$$R_{\alpha\gamma} = (1/2)(h_{\alpha\delta,\gamma}{}^\beta + h_{\beta\gamma,\alpha}{}^\beta - h_{\alpha\gamma,\beta}{}^\beta - h_\beta{}^\beta{}_{,\alpha\gamma}) + \mathrm{O}(h^2), \tag{7}$$

$$R = h_{\alpha\beta}{}^{,\alpha\beta} - h_\beta{}^\beta{}_{,\alpha}{}^\alpha + \mathrm{O}(h^2), \tag{8}$$

where $\mathrm{O}(h^2)$ denotes terms of order of $h_{\alpha\beta}h_{\mu\nu}$ or smaller. Now we choose the harmonic gauge condition for $h_{\alpha\beta}$,

$$[h_{\alpha\beta} - (1/2)\eta_{\alpha\beta}(\mathrm{Tr}\ h)]^{,\beta} = 0 + \mathrm{O}(h^2), \text{ i.e., } h_{\alpha\beta}{}^{,\beta} = (1/2)\ (\mathrm{Tr}\ h)_{,\alpha} + \mathrm{O}(h^2), \tag{9}$$

where $\mathrm{Tr}(h)$ is defined as the trace of $h_\alpha{}^\beta$, i.e. $\mathrm{Tr}(h) \equiv h_\alpha{}^\alpha$. Now the linearized Einstein equation can be derived from (3), (7) & (9) and written in the form:

$$h_{\mu\nu}{}^{,\beta}{}_\beta = -16\pi G_\mathrm{N}[T_{\mu\nu} - (1/2)(\eta_{\mu\nu}T)] + \mathrm{O}(h^2). \tag{10}$$

This is the linearized wave equation for GR. The corresponding equation for electromagnetism is

$$A_\mu{}^{,\beta}{}_\beta = 4\pi J_\mu, \tag{11}$$

with gauge condition

$$A_\alpha{}^{,\alpha} = 0. \tag{12}$$

The retarded solution of equation (12) is

$$A_\mu = \int (J_\mu/r)_\mathrm{retarded}\ (d^3x'). \tag{13}$$

Analogously, the solution of equation for GR in the harmonic gauge is

$$h_{\mu\nu} = -[(4G_\mathrm{N})/(c^4)]\ \int \{[T_{\mu\nu} - (1/2)\ g_{\mu\nu}T]/r\}_\mathrm{retarded}\ (d^3x') + \mathrm{O}(h^2). \tag{14}$$

In the linearized scheme, it is useful to represent the solution $h_{\mu\nu}(x, y, z, t)$ *outside of the source* region by its spectral components with wave vector $(k_x, k_y, k_z)$ and frequency $f$. First, find the Fourier transform $h_{\mu\nu}(k_x, k_y, k_z)$ of $h_{\mu\nu}(x, y, z, t)|_{t=0}$:

$^{(k)}h_{\mu\nu}(k_x, k_y, k_z) \equiv (c)^{-3} \int h_{\mu\nu}(x, y, z, 0) \exp(-\mathrm{i}k_x x - \mathrm{i}k_y y - \mathrm{i}k_z z) \, (dxdydz).$ (15)

The integration is from $-\infty$ to $\infty$ for each integration variable. From equation (10), for each spectral components, the frequency $f$ is given by the dispersion relation

$f = (c/(2\pi)) \, (k_x{}^2 + k_y{}^2 + k_z{}^2)^{1/2} \equiv (c/(2\pi)) \, k.$ (16)

Hence the solution is

$h_{\mu\nu}(x, y, z, t) = (c/2\pi)^3 \int {}^{(k)}h_{\mu\nu}(k_x, k_y, k_z) \exp(\mathrm{i}k_x x + \mathrm{i}k_y y + \mathrm{i}k_z z - 2\pi\mathrm{i}ft) \, (dk_x dk_y dk_z),$ (17)

with $f$ given by (16).

For plane GW $h_{\mu\nu}(n_x x + n_y y + n_z z - ct)$ propagating in the $(n_x, n_y, n_z)$ direction with $n_x{}^2 + n_y{}^2 + n_z{}^2 = 1$, letting

$U \equiv u - ct \equiv n_x x + n_y y + n_z z - ct,$ (18)

we can resolve the plane GW into the following spectral representation:

$h_{\mu\nu}(u, t) \equiv h_{\mu\nu}(u - ct) = h_{\mu\nu}(U) = (c/2\pi) \int_{-\infty}^{\infty} {}^{(k)}h_{\mu\nu}(k) \exp(\mathrm{i}ku) \exp(-2\pi\mathrm{i}ft) \, (dk),$ (19)

with

$^{(k)}h_{\mu\nu}(k) \equiv (c)^{-1} \int_{-\infty}^{\infty} h_{\mu\nu}(U) \exp(-\mathrm{i}kU) \, (dU).$ (20)

The plane wave (19-20) can also be written as

$h_{\mu\nu}(u, t) \equiv h_{\mu\nu}(u - ct) = h_{\mu\nu}(U) = \int_{-\infty}^{\infty} {}^{(f)}h_{\mu\nu}(f) \exp(2\pi\mathrm{i}fU/c) \, (df),$ (21)

with

$^{(f)}h_{\mu\nu}(f) \equiv {}^{(k)}h_{\mu\nu}(k=2\pi f/c) = \int_{-\infty}^{\infty} h_{\mu\nu}(u - ct)|_{u=0} \exp(2\pi\mathrm{i}ft) \, (dt).$ (22)

We note that since $h_{\mu\nu}(t)$ is real, $^{(f)}h_{\mu\nu}(-f) = {}^{(f)}h_{\mu\nu}*(f)$ and

$h_{\mu\nu}(U) = \int_0^\infty 2|{}^{(f)}h_{\mu\nu}(f)| \cos(2\pi fU/c) \, (df) = \int_0^\infty 2f \, |{}^{(f)}h_{\mu\nu}(f)| \cos(2\pi fU/c) \, d(ln \, f).$ (23)

From (21, 22) and the Parseval's equality, we have

$\int_{-\infty}^{\infty} |h_{\mu\nu}(t)|^2 \, (dt) = \int_{-\infty}^{\infty} |{}^{(f)}h_{\mu\nu}(f)|^2 \, (df).$     (No summations in the indices $\mu$ and $\nu$) (24)

The squared-amplitude integral is equal to its squared-spectral-amplitude integral. One can also obtain a similar identity relating the integral on the (absolute) square of $h_{\mu\nu}(x, y, z, t)$ over $(x, y, z)$ and the integral on the absolute square of $^{(k)}h_{\mu\nu}(k_x, k_y, k_z)$ over $(k_x, k_y, k_z)$ using (15), (17) and the Parseval's equality in 3 dimensions.

For weak GW $h_{\mu\nu}$ propagating in the space-time background $g_{\mu\nu}$ (i.e., the total space-time metric is $g_{\mu\nu} + h_{\mu\nu}$.), Isaacson[20,21] showed that the GW stress-energy averaged over several wavelength is

$$t_{\mu\nu} = (c^4/32\pi G_N) <\partial_\mu h^{TT}{}_{\alpha\beta} \; \partial_\nu h^{TT\alpha\beta}>. \tag{25}$$

Here $h^{TT}{}_{\mu\nu}$ is the transverse traceless part of $h_{\mu\nu}$. *In the special harmonic gauge called radiation gauge (similar to radiation gauge in electrodynamics)*, $h^{TT}{}_{\mu\nu} = h_{\mu\nu}$. Far from the sources, the GW can be approximated by plane waves. For a wave propagating in the $z$-direction, the only non-vanishing components of $h_{\mu\nu}$ in radiation gauge are $h_{11}$, $h_{22}$ ($= -h_{11}$), $h_{12}$ and $h_{21}$ ($= h_{12}$). The mass-energy density $t_{00}$ and mass-energy flux $ct_{03}$ are given by

$$\rho c^2 = t_{00} = t_{03} = (c^4/16\pi G_N) <[(1/4) \; (\partial_0 h_{11} - \partial_0 h_{22})^2 + (\partial_0 h_{12})^2]>$$
$$= (c^2/16\pi G_N) <(\partial_0 h_+{}^2 + \partial_0 h_\times{}^2)>, \tag{26}$$

in agreement with Ref. [22]. Here $h_+$ ($\equiv (1/2)(h_{11} - h_{22}) = h_{11} = -h_{22}$) is the amplitude of $e_1$-polarization (+-polarization); $h_\times$ ($\equiv h_{12} = h_{21}$) is the amplitude of $e_2$-polarization (×-polarization). Due to gauge (coordinate) invariance from the linearized wave equation (10) in GR, for plane GW waves in the direction of z-axis, there are two polarizations $\boldsymbol{e}_+$ and $\boldsymbol{e}_\times$:

$$\boldsymbol{e}_+ = (\underline{x}\,\underline{x} - \underline{y}\,\underline{y}), \; \boldsymbol{e}_\times = (\underline{x}\,\underline{y} + \underline{y}\,\underline{x}), \tag{27}$$

with $\underline{x}$ and $\underline{y}$ the unit vectors in the directions of x-axis and y-axis. The product $\underline{x}\,\underline{x}$ is tensor product. The metric tensor of $\boldsymbol{e}_+$-polarization GW is $h_+\boldsymbol{e}_+$; that of $\boldsymbol{e}_\times$-polarization GW is $h_\times\boldsymbol{e}_\times$. The total GW metric $\boldsymbol{h}$ is

$$\boldsymbol{h} = h_+\boldsymbol{e}_+ + h_\times\boldsymbol{e}_\times; \text{ in component form, } h_{\mu\nu} = h_+ e_{+\mu\nu} + h_\times e_{\times\mu\nu}. \tag{28}$$

GW with $e_+$-polarization contributes to the first term of the energy density formula (26) with squared amplitude $(\partial_0 h_+)^2$; GW with $e_\times$-polarization contributes to the second term of the energy density formula (26) with squared amplitude $(\partial_0 h_\times)^2$.

Far from the GW sources as it is in the present experimental/observational situations, the plane wave approximation is valid. Space averages can be replaced with time averages. For orthogonal modes, the energy can be added in quadrature. For multi-frequency plane GW, the total energy density in the spectral representation of (26) becomes then

$$\rho c^2 = t_{00} = (c^2/16\pi G_N) \int_{-\infty}^{\infty} (2\pi)^2 f^2 \; [|^{(f)}h_+(f)|^2 + |^{(f)}h_\times(f)|^2] \; (df) \equiv \int_0^{\infty} {}^{(E)}S_h(f) \; df. \tag{29}$$

$^{(E)}S_h(f)$ is defined as the (one-sided) energy spectral density of $h$ and is given by

$${}^{(E)}S_h(f) = (\pi c^2/2G_N) f^2 \; [|^{(f)}h_+(f)|^2 + |^{(f)}h_\times(f)|^2] = (\pi c^2/8G_N) \; f \, S_h(f), \tag{30}$$

with

$$S_h(f) = 4 f \, (|^{(f)}h_+(f)|^2 + |^{(f)}h_\times(f)|^2) = f^{-1} \; (h_c(f))^2;$$
$$S_{hA}(f) = 4 f \, |^{(f)}h_A(f)|^2 = f^{-1} \; (h_{cA}(f))^2 \text{ for a single polarization A (A = +, ×)}, \tag{31}$$

the spectral power density of $h$ and $h_A$ respectively and $h_c(f) \equiv 2 f [(|{}^{(f)}h_+(f)|^2 + |{}^{(f)}h_\times(f)|^2)]^{1/2}$; $h_{cA}(f) \equiv 2 f |{}^{(f)}h_A(f)|$, (31a)

the characteristic strains. For unpolarized GWs, $|{}^{(f)}h_+(f)|^2 = |{}^{(f)}h_\times(f)|^2$ and we have

$${}^{(E)}S_h(f) = (\pi c^2/G_N) f^2 |{}^{(f)}h_+(f)|^2 = (\pi c^2/G_N) f^2 |{}^{(f)}h_\times(f)|^2 = (\pi c^2/4G_N) f S_{hA}(f). \quad (32)$$

From (30), the energy density is proportional to $h_A{}^2$ for a particular polarization. General GW can be resolved into superposition of plane GWs, the formula (29-32) are still applicable. For an early motivation and an in-step mathematical derivation, see, e.g. [23] and [24] respectively.

For background or foreground stochastic GWs, it is common to use the critical density $\rho_c$ for closing the universe as fiducial:

$$\rho_c = 3H_0{}^2/(8\pi G_N) = 1.878 \times 10^{-29} \text{ g/cm}^3, \quad (33)$$

where $H_0$ is the Hubble constant at present. Throughout this article, we use the Planck 2015 value 67.8 ($\pm$ 0.9) km s$^{-1}$ Mpc$^{-1}$ for $H_0$ [25]. In the cosmological context, it is more convenient to define a *normalized GW spectral energy density* $\Omega_g(f)$ and express the GW spectral energy density in terms of *the energy density per logarithmic frequency interval divided by the cosmic closure density $\rho_c$* for a cosmic GW sources or background, i.e.,

$$\Omega_{gw}(f) = (f/\rho_c) \, d\rho(f)/df = (\pi/8G_N) f^3 S_h(f)/(3H_0{}^2/(8\pi G_N)) = (\pi^2/3H_0{}^2) f^3 S_h(f)$$
$$(=(2\pi^2/3H_0{}^2) f^3 S_{hA} (f) \text{ for unpolarized GW}). \quad (34)$$

For the very-low-frequency band, the ultra-low-frequency band and the extremely-low-frequency band, this is a common choice.

From equation (14), one can derive the quadrupole formulas of the gravitational radiation metric and the radiated power at the lowest approximation:[22]

$$h_{ij}(t, x, y, z) = -[(2G_N/(c^6 r)) \, d^2Q_{ij}/dt^2]_{\text{retarded}}; \quad (35)$$

$$dP/d\Omega = [G_N/(8\pi c^5)] [(d^3Q_{ij}/dt^3) e_{ij}]^2. \quad (36)$$

Here $dP/d\Omega$ is the power radiated into the solid angle $d\Omega$ in the polarization $e_{ij}$, $Q_{ij}$ ($= \int \rho \, x_i \, x_j \, d^3x$) is the moment of inertia of the radiating system, and $e_{ij}$ is the polarization of the emitted GW. Summed over two polarizations and integrated over solid angles, the total power emitted is

$$P = [G_N/(5c^5)] [(d^3Q_{ij}/dt^3)(d^3Q_{ij}/dt^3) - (1/3)(d^3Q_{ii}/dt^3) \, d^3Q_{jj}/dt^3)]. \quad (37)$$

Inserting the moment of inertia of the binary Keplerian orbit motion into (46) and average over one orbit period, Peters and Mathews[26] obtained the following formula for the gravitational radiation loss:

$$< P > = [32G_N{}^4/(5c^5)] [M_1{}^2 M_2{}^2 (M_1 + M_2)/a^5(1 - e)^{7/2}] [1 + (73/24) e^2 + (37/96) e^4]. \quad (38)$$

Here $M_1$ and $M_2$ are the two masses of the binary, $e$ is the eccentricity of the elliptic orbit, and $a$ the semi-major axis. Peters[27] further obtained the average angular momentum emission rate:

$$<dL/dt> = -[32G_N^{7/2}/(5c^5)] [M_1^2 M_2^2 (M_1 + M_2)^{1/2} / a^{7/2} (1 - e^2)^2] [1 + (7/8) e^2].\qquad(39)$$

From the Peters-Mathews radiation formula (47) and Peters' angular momenta radiation formula (48), the orbital period $P_b$ decay rate can be calculated as[27]

$$dP_b/dt = -(192\pi/5) (P_b/2\pi)^{-5/3} [1 + (73/24) e^2 + (37/96) e^4] (1 - e)^{7/2} M_1 M_2 (M_1 + M_2)^{-1/3}$$
$$(40)$$

From (39, 40) Peters obtained the time evolution equations for $<da/dt>$ and $<de/dt>$, and found the time dependence of the semi-major axis $a(t)$ and the merging time $T_c(a_0)$ for circular orbits starting from initial semi-major axis $a = a_0$:

$$a(t) = (a_0^4 - 4\beta t)^{1/4}; T_c(a_0) = a_0^4/(4\beta), \text{ with } \beta \equiv [64G_N^3/(5c^5)] M_1 M_2 (M_1 + M_2),\qquad(41)$$

in reasonable agreement with estimates from higher-order approximations and results from numerical relativity.

For a binary system of masses $M_1$ and $M_2$ with Schwarzschild radius $R_1$ and $R_2$, the strain $h$ calculated from (35) of its emitted gravitational radiation is of the order of

$$h \approx R_1 R_2 / Dd ,\qquad(42)$$

where $d$ is the distance between $M_1$ and $M_2$, $D$ the distance to the observer. For neutron star or black hole, $d$ can be of the order of Schwarzschild radius, and the estimation can be simplified:

$$h \leq R/D.\qquad(43)$$

For black hole of solar masses, $R = 3$ km, and $d = 10^8$ l.y., $h \leq 3 \times 10^{-21}$; for inspiral of neutron star binaries, the GW strain generated is smaller.

GWs in GR have 2 independent polarizations. GWs in a general metric theory of gravity can have up to 6 independent polarizations according to the Riemann tensor classification of Eardley *et al.* [28]; in terms of helicity, there are two scalar modes, one helicity $+2$ mode, one helicity $+1$ mode, one helicity $-1$ mode and one helicity $-2$ mode [29]. Therefore by measuring the GW polarizations, different theories can be distinguished and tested. For a general metric theory with additional fields (scalar, vector, etc.), there are monopole and/or dipole contributions to the quadrupole radiation formula (36). However, due to conservation of mass and conservation of linear momentum in the Newtonian order, the leading order of monopole and dipole contributions are of the same order or less compared with the quadrupole contribution in GR [30]. Nevertheless, experiments/observations do distinguish them. For example, pulse timing observations on the relativistic pulsar-white dwarf binary PSR J1738+0333 have given stringent tests on some of these theories already [8].

### 3. Methods of GW detection, and their sensitivities

Similar to the frequency classification of electromagnetic waves to radio wave, millimeter wave, infrared, optical, ultraviolet, X-ray and γ-ray etc., in Table 1, we have compiled a complete frequency classification of GWs. This classification together with the current and aimed sensitivities of various detection methods plus predicted GW source strengths are plotted in Fig. 1-4. Fig. 1 shows the spectrum classification together with detection methods and projects. Fig. 2-4 show respectively the characteristic strain $h_c$ versus frequency plot, the strain psd (power spectral density) amplitude $[S_h(f)]^{1/2}$ versus frequency plot and the normalized GW spectral energy density $\Omega_g$ versus frequency plots for various GW detectors and sources. Detailed accounts and explanations of Fig. 2-4 are given in the following subsections and in Sec. 4.

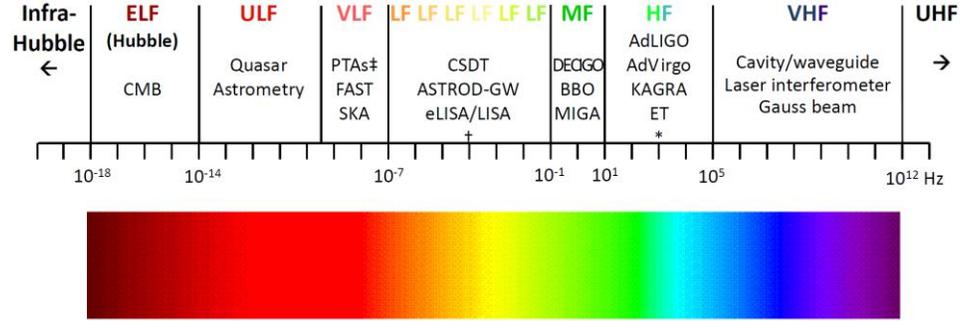

* AIGO, AURIGA, EXPLORER, GEO, NAUTILUS, MiniGRAIL, Schenberg.

† OMEGA, gLISA/GEOGRAWI, GADFLI, TIANQIN, ASTROD-EM, LAGRANGE, ALIA, ALIA-descope.

‡ EPTA, NANOGrav, PPTA, IPTA.

Fig. 1. The Gravitation-Wave (GW) Spectrum Classification (updated from [16, 17]).

For the methods of detecting GWs, we first classify them into real-time detection and imprint detection. For real-time detection, we use the time scale of 100 yr – the life span of a human being. Although this scale could be extended, it is at least good for next few hundred years. Above 300 pHz [∼(100 yr)$^{-1}$], real-time detections are possible. These detections include using resonators, interferometers and pulsar timing for detection in the first six GW bands in Table 1. Below 300 pHz, the detections are possible on GW imprints. Imprint (or snapshot) detections include (i) using the method of quasar astrometry for detection in the ultralow frequency GW band, (ii) using CMB observations for detection in Hubble frequency GW band, and (iii) using indirect verifications of primordial (inflationary or noninflationary) cosmological models in the beyond the Hubble frequency band.

There are basically two kinds of GW detectors for real-time detection – (i) the resonant type: GW induces resonances in detectors (metallic bars, metallic spheres, resonant cavities…) to enhance sensitivities; (ii) detectors measuring distance change using microwave/laser/X-ray/atom/molecule… between/among suspended/floating test bodies. In the case of Pulsar Timing Arrays (PTAs) for detection in the very low frequency GW band, the floating test bodies are the pulsars and observatories while the relative distance change are through pulsar timing variations. Two crucial issues in real-time GW detection are (i) to lower disturbance effects and/or to model the residuals: suspension isolation, drag-free to decrease the effects of surrounding disturbances, and appropriate modeling of the motion and the disturbances; (ii) to increase measurement sensitivity: capacitive sensing, microwave sensing, SQUID transducing, optical sensing, X-ray sensing, atom sensing, molecule sensing and timing….

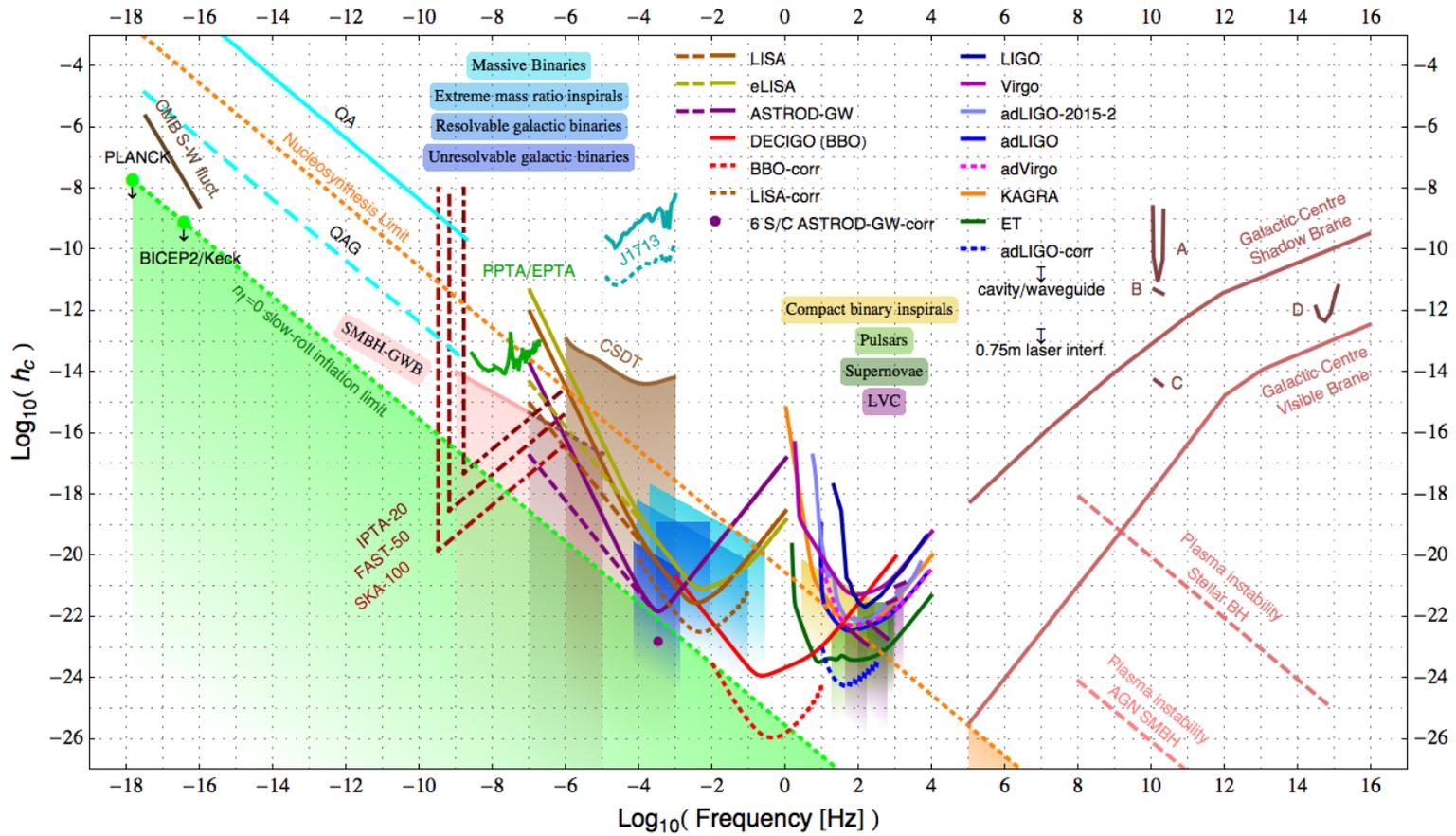

Fig. 2. Characteristic strain $h_c$ vs. frequency for various GW detectors and sources. [QA: Quasar Astrometry; QAG: Quasar Astrometry Goal; LVC: LIGO-Virgo Constraints; CSDT: Cassini Spacecraft Doppler Tracking; SMBH-GWB: Supermassive Black Hole-GW Background.]

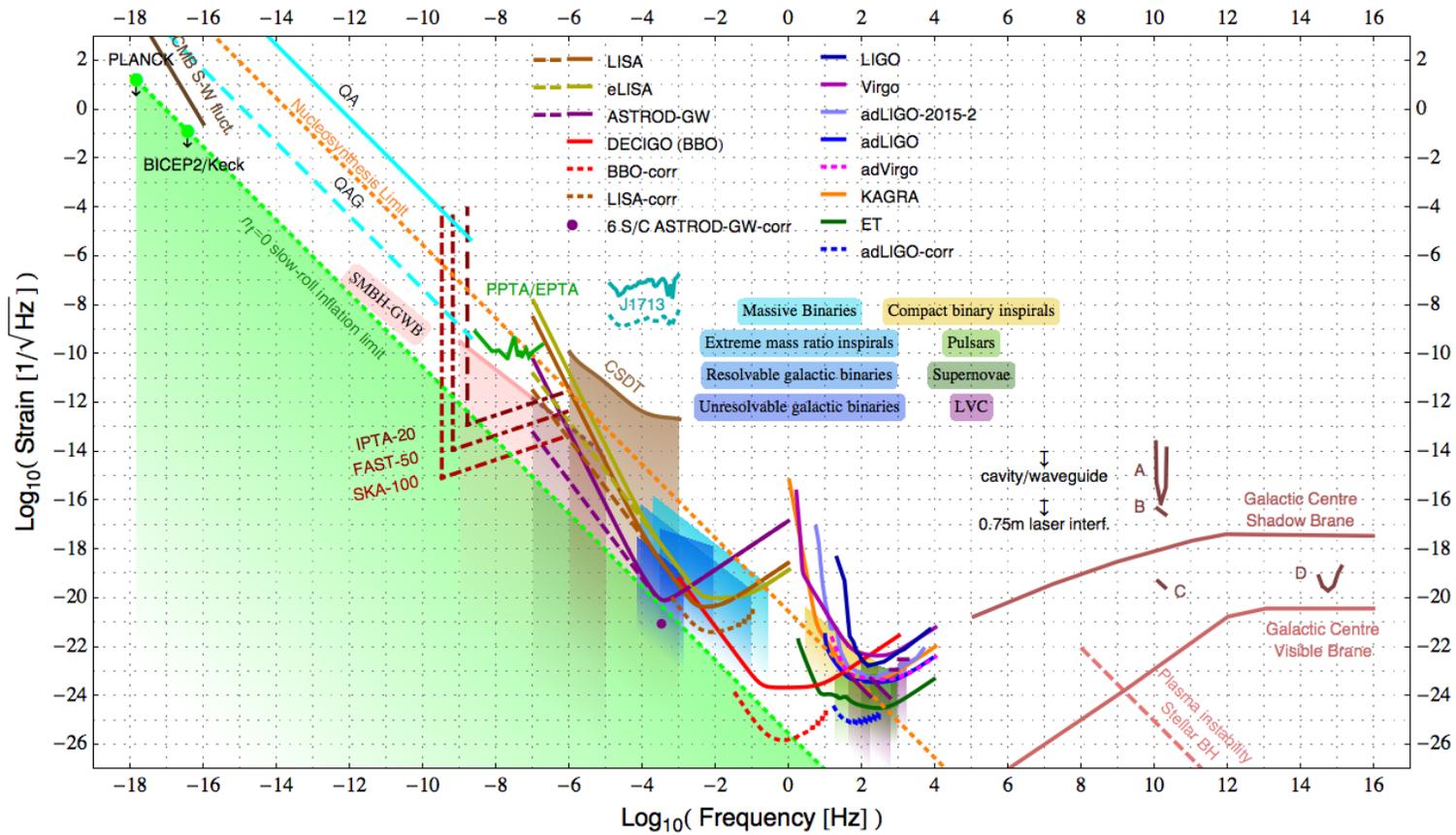

Fig. 3. Strain power spectral density (psd) amplitude vs. frequency for various GW detectors and GW sources. See Fig. 2 caption for the meaning of various acronyms.

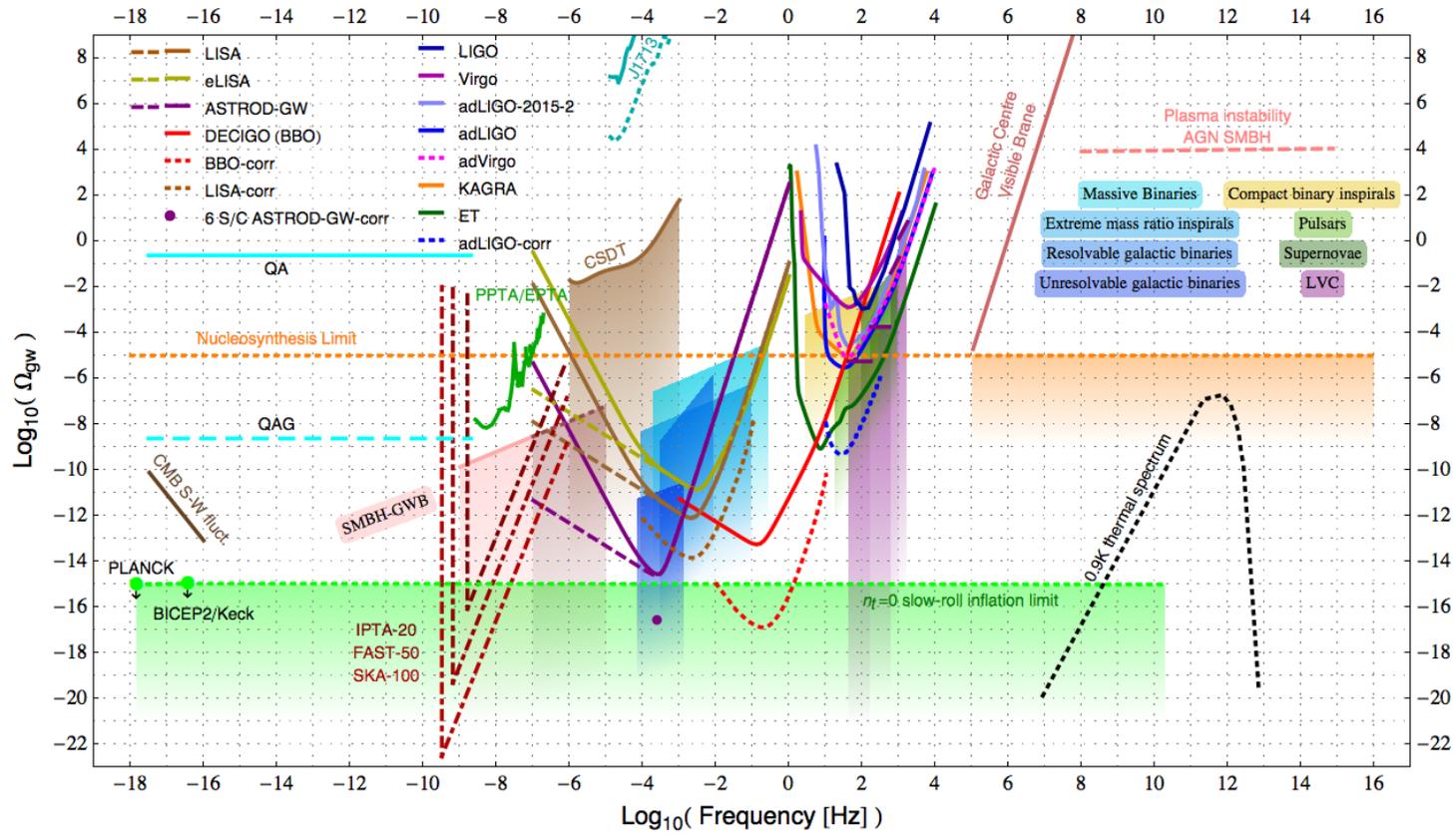

Fig. 4. Normalized GW spectral energy density $\Omega_{gw}$ vs. frequency for GW detector sensitivities and GW sources. See Fig. 2 caption for the meaning of various acronyms.

### 3.1. Sensitivities

The input and output of a detector are scalar quantities. The input of a GW detector is a time series $h(t)$ of GW signals which can be written as a functional of the GW metric $h_{\alpha\beta}(\underline{x}, \underline{t})$. For weak GW as in most situations, this functional can be linearized and approximated by a linear functional $D$ of $h_{\alpha\beta}(\underline{x}, \underline{t})$:

$$h(t) = D(h_{\alpha\beta}(\underline{x}, \underline{t})). \tag{44}$$

For a stationary local detector, $D$ may further be reduced to a constant tensor $D^{\alpha\beta}$ such that

$$h(t) = D^{\alpha\beta} h_{\alpha\beta}(\boldsymbol{x}, t). \tag{45}$$

In a transverse, traceless coordinate gauge, $h(t)$ is further reduced to

$$h(t) = D^{ij} h^{\text{TT}}{}_{ij}(\boldsymbol{x}, t). \tag{46}$$

$D^{ij}$ (or $D^{\alpha\beta}$) is called the detector tensor which depends on detection geometry. As an example, for a GW interferometer oriented with two arms on the x-axis and y-axis with nearly equal arm lengths in the long wavelength limit, the detector tensor has $D^{11} = 1/2$, $D^{22} = -1/2$ and all other components vanishing; we have $h(t) = h_{11}(t)$, $^{(\text{f})}h(f) = {}^{(\text{f})}h_{11}(f) = {}^{(\text{f})}h_{+}(f)$ and $h_{c+}(f) \equiv 2f \; |^{(\text{f})}h_{+}(f)|$.

In the case of linear response, the detector output $^{(\text{f})}h^{(\text{out})}(f)$ is related to the input by

$$^{(\text{f})}h^{(\text{out})}(f) = T(f) \times {}^{(\text{f})}h(f), \quad (\text{or simply } h^{(\text{out})}(f) = T(f) \times h(f) \text{ without heavy notations}) \tag{47}$$

where $T(f)$ is the transfer function or the response function of the detector. In the output of any detector there will be noise also. The total output $s^{(\text{out})}(t)$ is the addition of the GW signal output $h^{(\text{out})}(t)$ and the noise output $n^{(\text{out})}(t)$:

$$s^{(\text{out})}(t) = h^{(\text{out})}(t) + n^{(\text{out})}(t); \tag{48}$$

in frequency space the total output $^{(\text{f})}s^{(\text{out})}(f)$ is

$$^{(\text{f})}s^{(\text{out})}(f) = {}^{(\text{f})}h^{(\text{out})}(f) + {}^{(\text{f})}n^{(\text{out})}(f). \tag{49}$$

Habitually, it is convenient to refer and compare noise at the input port by defining

$$^{(\text{f})}n(f) \equiv [T(f)]^{-1} \times {}^{(\text{f})}n^{(\text{out})}(f). \tag{50}$$

From (50) we have

$$^{(\text{f})}s^{(\text{out})}(f) = [T(f)] \times [{}^{(\text{f})}h(f) + {}^{(\text{f})}n(f)], \tag{51}$$

And we can define

$$^{(\text{f})}s(f) \equiv [T(f)]^{-1} \times {}^{(\text{f})}s^{(\text{out})}(f) = {}^{(\text{f})}h(f) + {}^{(\text{f})}n(f) \tag{52}$$

to be the total input signal. In time domain, we have then

$$s(t) = h(t) + n(t). \tag{53}$$

It is convenient to take $n(t)$ as the detector noise.

It is also convenient and practical to assume that the detector noise is stationary and Gaussian, and the different Fourier components are independent. We then have

$$<{}^{(f)}n{*}(f)\,{}^{(f)}n(f')> = (1/2)\delta(f-f')S_n(f) \tag{54}$$

where $S_n(f)$ is defined by the equation. From this equation, one can derive

$$<n^2(t)> = <n^2(t=0)> = \int_{-\infty}^{\infty}(df)(df')<{}^{(f)}n{*}(f)\,{}^{(f)}n(f')> = \int_0^{\infty}(df)\,S_n(f). \tag{55}$$

Hence $S_n(f)$ is called the noise power spectrum, the noise power spectrum density (noise psd), the noise spectral density, or the noise spectral sensitivity. It is one-sided since the integration only takes on the positive axis. For a more detailed derivation of (64), we refer the reader to ref.'s [24, 31]. For a dimensionless description of noise power at a particular frequency, one usually use noise amplitude $h_n(f)$ which is defined as

$$h_n(f) \equiv [f\,S_n(f)]^{1/2}. \tag{56}$$

For comparison, we note that for a GW interferometer oriented with two arms oriented on the x-axis and y-axis with nearly equal arm lengths in the long wavelength limit, the GW signal $h(t) = h_{11}(t) = h_{+}(t)$ (${}^{(f)}h(f) = {}^{(f)}h_{11}(f) = {}^{(f)}h_{+}(f)$) with GW propagating perpendicular to xy-plane corresponds to detector geometry with $D_{\mu\nu} = e_{+\mu\nu}$. Its associated strain psd $S_h(f)$ is

$$S_h(f) = 4f\,|{}^{(f)}h(f)|^2 = |2\,f^{1/2}\,{}^{(f)}h_{11}(f)|^2 = |2\,f^{1/2}\,{}^{(f)}h_{+}(f)|^2 = |f^{-1/2}\,{}^{(f)}h_{c+}(f)|^2 = |f^{-1/2}\,{}^{(f)}h_c(f)|^2. \tag{57}$$

In general, GW detector has different geometric sensitivity to monochromatic GW coming from different directions and with different polarization. Hence each detector has its own pattern function of directions and polarizations. In plotting GW sensitivities, one usually takes average over directions and polarizations for a detector.

In the discussion of GW sensitivities and GW signal strengths, there are three customary ways to plot: characteristic strain $h_c(f)$ vs. frequency $f$, square-root power spectral density $[S_h(f)]^{1/2}$ vs. frequency $f$, and the normalized GW spectral energy density $\Omega_g$ vs. frequency $f$. From (57), for this rather general case, we define the (dimensionless) characteristic strain for the singal ${}^{(f)}h(f)$ as in (31a)

$$h_c(f) \equiv 2f\,|{}^{(f)}h_{11}(f)| = 2f\,|{}^{(f)}h_{+}(f)|. \tag{58}$$

With this definition, we have from (57) and (58)

$$S_h(f) = f^{-1}\,|h_c(f)|^2 = f\,|2\,{}^{(f)}h_{11}(f)|^2 = f\,|2\,{}^{(f)}h_{+}(f)|^2. \tag{59}$$

Now we relate the three quantities $h_c(f)$, $[S_h(f)]^{1/2}$ and $\Omega_{gw}(f)$ using equations (34) and (58):

$$\Omega_{gw}(f) = (2\pi^2/3H_0^2)\, f^3\, S_h(f); \; h_c(f) = f^{1/2}\, [S_h(f)]^{1/2}. \qquad (60)$$

Table 2 compiles the conversion factors among the characteristic strain $h_c(f)$, the strain psd (power spectral density) $[S_h(f)]^{1/2}$ and the normalized spectral energy density $\Omega_g(f)$. In using (60) and Table 2, especially the conversion to $\Omega_{gw}(f)$, we assume a baseline detector and source configuration just mentioned. For other configuration, its specific detector geometry, source geometry and GW polarization need to be taken care of.

Table 2. Conversion factors among the characteristic strain $h_c(f)$, the strain psd (power spectral density) $[S_h(f)]^{1/2}$ and the normalized spectral energy density $\Omega_{gw}(f)$.

|  | Characteristic strain $h_c(f)$ | Strain psd $[S_h(f)]^{1/2}$ | Normalized spectral energy density $\Omega_{gw}(f)$ |
|---|---|---|---|
| $h_c(f)$ | $h_c(f)$ | $f^{1/2}\,[S_h(f)]^{1/2}$ | $[(3H_0^2/2\pi^2 f^2)\Omega_{gw}(f)]^{1/2}$ |
| Strain psd $[S_h(f)]^{1/2}$ | $f^{-1/2}h_c(f)$ | $[S_h(f)]^{1/2}$ | $[(3H_0^2/2\pi^2 f^3)\Omega_{gw}(f)]^{1/2}$ |
| $\Omega_{gw}(f)$ | $(2\pi^2/3H_0^2)\,f^2\,h_c^2(f)$ | $(2\pi^2/3H_0^2)\,f^3\,S_h(f)$ | $\Omega_{gw}(f)$ |

In data analysis, the optimal signal to noise ratio $\eta$ that can be obtained is by using Wiener matched filter $^{(f)}h(f) / S_n(f)$:

$$\eta^2 = \int_0^\infty df\, [4\, |^{(f)}h(f)|^2 / S_n(f)]. \qquad (61)$$

Using equations (58) and (56), (61) can be written as

$$\eta^2 = \int_0^\infty df\, f^{-1}[h_c(f) / h_n(f)]^2 = \int_{-\infty}^\infty d(\log f)\, [h_c(f) / h_n(f)]^2. \qquad (62)$$

Hence in the log-log plot of characteristic strain vs. frequency, the square of signal to noise ratio is equal to the integral of the square of the ratio of characteristic strain of source over the characteristic noise strain. For large signal to noise ratio, it is approximately equal to the area between the characteristic strain curve and the characteristic noise strain curve in the detection bandwidth.

For the parameter fitting, the more discernable the structure, the better are the parameters fitted. Please see [32-35] for good accounts.

### 3.2. Very high frequency band (100 kHz − 1 THz) and ultrahigh frequency band (above 1 THz)

In the very high frequency band (100 kHz − 1 THz), there are two experiments completed. A cavity/waveguide detector, where the polarization of electromagnetic wave changes its direction under incoming gravitational wave, was operated at 100 MHz and gave upper limit of the background gravitational wave radiation of around $10^{-14}$ Hz$^{-1/2}$.[36] And a 0.75 m arm length laser interferometer, where synchronous amplification of the phase shift due to gravitational wave occurs, achieved a noise level limiting the existence of 100 MHz background gravitational wave down to $10^{-16}$ Hz$^{-1/2}$.[37,38] These two upper limits are marked on Fig. 3 and the corresponding limits on Fig. 2 and Fig. 4.

Cruise has described two types of magnetic conversion prototype detectors A and D being commissioned at Birmingham in Sec. 9 of [39]. The basic principle of a magnetic conversion detector is to convert GWs in a laboratory magnetic field to electromagnetic waves which are then focused or concentrated on one detector element to be measured.. Detector A has a room temperature microwave receiver sensing the waveguide conversion volume with a magnetic field 0.2 T. The expected sensitivity curve of prototype detector A for one year integration with 1 MHz bandwidth is shown in Fig. 1 of [39] as curve A. In the same figure, curve B shows the expected sensitivity for a larger detector, $60 \times 500 \times 800$ mm$^3$ having a 20 K noise temperature amplifier and curve C shows that for a pair of such detectors in correlation over a one year period. Curve D is for the detector with the cooled CCD sensing the same waveguide and field as A, also for a one year integration.

We put curves A, B, C and D from Fig. 1 of [39] on Fig. 3 in this section. The definition of $\Omega_{gw}(f)$ in Eq. (1) of Ref. [39] is the same as ours; while their conversion to their $h$ (Eq. (2) of ref. [39] is

$$h = 5.8 \times 10^{-22} \ (100/f)^{3/2} \ (\Omega_{gw}(f))^{1/2}. \tag{63}$$

Our strain psd $[S_h(f)]^{1/2}$ using Planck $H_0$ [25] is

$$[S_h(f)]^{1/2} = [(3H_0^2/2\pi^2 f^3)\Omega_{gw}(f)]^{1/2} = 8.57 \times 10^{-22} \ (100/f)^{3/2} \ (\Omega_{gw}(f))^{1/2}. \tag{64}$$

Therefore their $h$ is basically our $[S_h(f)]^{1/2}$ with a multiplicative factor 0.677. Hence Fig. 1 of [39] corresponds to our Fig. 3 basically. We adjust the factor 0.677 for the nucleosynthesis limit in Fig. 4 and corresponding places in Fig. 2 and Fig. 3. We have not adjusted other parts of Fig. 1of [39] while transport to our figures since the multiplicative factor is not large in our log-log plots.

Curve C and curve D have sensitivities in strain psd $[S_h(f)]^{1/2}$ close to $10^{-20}$ for frequencies around $10^{10}$ Hz in the very high frequency band and $10^{15}$ Hz in the ultrahigh frequency band respectively. The corresponding curves are also shown in Fig. 2 and Fig. 4. Possible sensitivity enhancements have been suggested by generating electromagnetic power depending linearly on the GW amplitudes;[40] however, the associated noise issues are still pending on solutions.[41,39] Signal amplitudes from various GW sources are summarized in Sec. 4.6.

### 3.3. High frequency band (10 Hz – 100 kHz)

Most of the current activities of gravitational wave detection on the ground or in the underground are in the high frequency band. In the following, we summarize the activities and sensitivities. For a detailed exposition, we refer to [42].

In this band, the cryogenic resonant bar detectors have already reached a strain spectral sensitivity of $10^{-21}$ Hz$^{-1/2}$ in the kHz region. NAUTILUS put an upper limit on periodic sources ranging from $3.4 \times 10^{-23}$ to $1.3 \times 10^{-22}$ depending on frequency in their all-sky search.[43] The AURIGA-EXPLORER-NAUTILUS-Virgo Collaboration applied a methodology to the search for coincident burst excitations over a 24 h long joint data set.[44] The MiniGRAIL[45] and Schenberg[46] cryogenic spherical GW detectors are for omnidirectional GW detection.

Major detection efforts in the high frequency band are in the long arm laser interferometers. The TAMA 300 m arm length interferometer,[47] the GEO 600 m interferometer,[48] and the kilometer size laser-interferometric GW detectors --- LIGO[49] (two 4 km arm length, one 2 km arm length) and VIRGO[50] all achieved their original sensitivity goals basically. Around the frequency 100 Hz, the LIGO and Virgo sensitivities are both in the level of $10^{-23}$ (Hz)$^{-1/2}$. The LIGO and Virgo achieved sensitivity curves are shown in Fig.'s 2-4.[49,50] Interference spikes are taken out for clarity in the presentation in these figures. Various limits on the GW strains for different sources become significant. For example, analyses of data from S6 (sixth science run) of LIGO and GEO600 gravitational wave detectors and VSR 2 and VSR 4 (Virgo science runs) of Virgo detector set strain upper limits on the gravitational wave emission from 195 radio pulsars; specifically, the strain upper limit on the Vela pulsar is comparable to the spin-down limit and that on the Crab pulsar is about a factor of 2 below the spin-down limit.[51] The 2009 analysis of the data from a LIGO two-year science run constrained the normalized spectral energy density $\Omega_{gw}(f)$ of the stochastic GW background in the frequency band around 100 Hz, to be $6.9 \times 10^{-6}$ at 95% confidence.[52] This search for the stochastic background improved on the indirect limit from the Big Bang nucleosynthesis at 100 Hz. In 2014 further improvement and refinement on the limit of the stochastic GW background were obtained from the analysis of the 2009–2010 LIGO and Virgo Data.[53] Assuming a stochastic GW spectrum of

$$\Omega_{gw}(f) = \Omega_\alpha (f/f_{ref})^\alpha, \tag{65}$$

LIGO and Virgo collaboration placed 95% confidence level upper limits on the normalized spectral energy density of the background in each of four frequency bands spanning 41.5–1726 Hz:

$\Omega_{gw}(f) < 5.6 \times 10^{-6}$, for the frequency band 41.5-169.25 Hz;
$\Omega_{gw}(f) < 1.8 \times 10^{-4}$, for the frequency band 170-600 Hz;
$\Omega_{gw}(f) < 0.14 \ (f/900 \text{ Hz})^3$, for the frequency band 600-1000 Hz;
$\Omega_{gw}(f) < 1.0 \ (f/900 \text{ Hz})^3$, for the frequency band 1000-1726 Hz. $\qquad$ (66)

These constraints [LVC: LIGO-Virgo Constraints] are plotted on Fig. 4 and the corresponding constraints on Fig. 2 and Fig. 3.

Also, in the analysis of jointly conducted science runs (LIGO S6 and Virgo VSR 2 and VSR 3), two kinds of search were done for possible GWs associated with 154 gamma ray bursts that were observed by satellite experiments in 2009-2010: the first search is for a signal from coalescence of two neutron stars or a neutron star and a black hole and the second search is for a burst-like gravitational wave signal from the collapse of a massive star. No signals were detected. This results places limits of 17 Mpc for no collapsing star, 16 Mpc for non-existence of the coalescence of binary neutron stars and 28 Mpc for that of a neutron star and a black hole associated with the observed 154 gamma ray bursts.[54]

Observations by all above long baseline laser interferometers have finished their first phase operation by 2010. Sensitivity improvement of one order of magnitude is underway by upgrading LIGO and Virgo as advanced interferometers, adLIGO[55] and adVirgo,[56] and by initiating a new project, KAGRA/LCGT.[57] This improvement will increase the detection volume by three orders of magnitudes. These GW detectors are the second-generation interferometers. The advanced LIGO is the earliest started and has achieved 3 times better sensitivity improvement already; it began its first observing run (O1) on September 18, 2015 searching for GWs. We plot the February-2015 achieved adLIGO sensitivity together with the planned strain sensitivities of adLIGO,[55] adVirgo[56] and KAGRA[57] on Fig.'s 2-4. KAGRA will be a cryogenic underground interferometer with 3 km arm length; it will already have some features of the third generation GW interferometers. ET (Einstein Telescope)[58] is a third generation GW interferometer. It will be a cryogenic underground interferometer with 10 km arm length. Its goal sensitivity is also plotted on Fig.'s 2-4.

As to the upper range of this band, it is noticed that every FSR (Free Spectral Range) relative to the lock point, there would be good sensitivity. The FSRs of LIGO and VIRGO/KAGRA are 37.5 MHz and 50 MHz. LIGO is considering/discussing this frequency. Although digitation under 100 kHz is not a technological feasibility problem, it is a practical problem in sampling/digitizing the data at these high frequencies. The upper range of the high frequency band is accessible to the km-sized GW interferometers.

### 3.4. Doppler tracking of spacecraft (1 μHz – 1 mHz in the low-frequency band)

Doppler tracking of spacecraft can be used to constrain (or detect) the level of low-frequency GWs.[59] The separated test masses of this GW detector are the Doppler tracking radio antenna on Earth and a distant spacecraft. Doppler tracking measures relative distance-change. Estabrook and Walquist analyzed[59] the effect of GWs passing through the line of sight of spacecraft on the Doppler tracking frequency measurements (see also [60]). From these measurements, GWs can be detected or constrained. The most recent measurements came from the Cassini spacecraft Doppler tracking (CSDT). Armstrong, Iess, Tortora, and Bertotti[61] used precision Doppler tracking of the Cassini spacecraft during its 2001–2002 solar opposition to derive improved observational limits on an isotropic background of low-frequency gravitational waves. They used the Cassini multilink radio system and an advanced tropospheric calibration system to remove the effects of leading noises — plasma and tropospheric scintillation to a level below the other noises. The resulting data were used to construct upper limits on the strength of an isotropic background in the 1 μHz to 1 mHz band.[61] The characteristic strain upper limit curve labelled CSDT in Fig. 2 is a smoothed version of the curve in the Fig. 4 of Ref. 61. The corresponding CSDT curves on the strain psd amplitude in Fig. 3 and the normalized spectral energy density in Fig. 4 are calculated using Table 2 for conversion. The minimal points on these curves are

$h_c(f) < 2 \times 10^{-15}$, at frequency about 0.3 mHz;

$[S_h(f)]^{1/2} < 8 \times 10^{-13}$, at several frequencies in the 0.2-0.7 mHz band;

$\Omega_{gw}(f) < 0.03$, at frequency 1.2 μHz. $\qquad\qquad$ (67)

The GW sensitivity of spacecraft Doppler tracking could still be improved by 1-2 order of magnitude with a space borne optical clock on board.[62]

The basic principle of spacecraft Doppler tracking, of spacecraft laser ranging, of space laser interferometers, and of pulsar timing arrays (PTAs) for GW detection are similar. In the development of further GW detection methods, Spacecraft Doppler tracking method has stimulated significant inspirations. ASTROD I (Astrodynamical Space Test of Relativity Using Optical Devices I)[63] using a space borne precision clock has included as one of its goals GW sensitivity improvement of the Cassini spacecraft Doppler tracking by one order of magnitude. The methods using space laser interferometers and using PTAs are two important methods of detecting GWs; their sensitivities will be discussed in Sec. 3.5 and Sec. 3.6 respectively.

### 3.5. Space interferometers (low-frequency band, 100 nHz – 100 mHz; middle-frequency band, 100 mHz – 10 Hz)

*Space laser interferometers for GW detection (eLISA/LISA,[64,65] ASTROD,[66,67] ASTROD-GW,[13-16,68,69] ASTROD-EM,[69,70] Super-ASTROD,[71] DECIGO,[72] Big Bang*

*Observer,*[73] *ALIA,*[74] *ALIA-descope,*[75] gLISA (*GEOGRAWI*),[76-78] *GADFLI,*[79] *LAGRANGE,*[80] *OMEGA,*[81] *and TIANQIN*[82]) hold the most promise with high signal-to-noise ratio. LISA[65] (Laser Interferometer Space Antenna) is aimed at detection of $10^{-4}$ to 1 Hz GWs with a strain sensitivity of $4 \times 10^{-21}/(Hz)^{1/2}$ at 1 mHz. There are abundant sources for eLISA/LISA, ASTROD, ASTROD-GW and Earth-orbiting missions: (i) In our Galaxy: galactic binaries (neutron stars, white dwarfs, etc.); (ii) Extra-galactic targets: supermassive black hole binaries, supermassive black hole formation; and (iii) Cosmic GW background. A date of launch of eLISA or substitute mission is set around 2034.[4]

Early in 2009, responding to the call for GW mission studies of CAS (Chinese Academy of Sciences), a dedicated GW mission concept ASTROD-GW with 3 S/C's orbiting near Sun-Earth Lagrange points L3, L4 and L5 respectively was proposed and studied. Before that, Super-ASTROD which was proposed in 1997[10] with S/C's in Jupiter-like orbits was studied as a dual mission for GW measurement and for cosmological model/relativistic gravity test.[71] With the proposal of ASTROD-GW, the baseline GW configuration of Super-ASTROD takes 3 S/C's orbiting respectively near Sun-Jupiter Lagrange points L3, L4 and L5. For the possibility of a down scaled version of ASTROD-GW mission, the ASTROD-EM with the orbits of 3 S/C's near Earth-Moon Lagrange points L3, L4 and L5 respectively has been under study.[70]

DECi-hertz Interferometer GW Observatory (DECIGO)[72] was proposed in 2001 with the aim of detecting GWs from early universe in the observation band (the middle frequency band) between the terrestrial band and the band of low-frequency space GW detectors. It will use a Fabry-Perot method (instead of a delay line method) as in the ground interferometers but with a 1000 km arm length. As a LISA follow-on, BBO (Big Bang Observer)[73] was proposed in the United States with a similar goal. A likely version of DECIGO/BBO is to have 12 S/Cs in LISA-like orbits with correlated detection. They will be used for the direct measurement of the stochastic GW background by correlation analysis.[83] 6S/C-ASTROD-GW has also been considered to possibly explore the relic GWs in the lower part of the low frequency band. ALIA[74] was proposed as a less-ambitious LISA follow-on. A de-scoped ALIA[75] has also proposed and under study.

After the end in 2011 of ESA-NASA partnership for flying LISA, NASA solicited "Concepts for the NASA Gravitational Wave Mission" proposals on September 27, 2011 for study of low cost GW missions (http://nspires.nasaprs.com/external/). gLISA[76] (geosynchronous LISA) and LAGRANGE[80] (Laser Gravitational-wave Antenna at Geo-lunar Lagrange points) was proposed and OMEGA[81] (Orbiting Medium Explorer for Gravitational Astronomy) re-emerged. OMEGA was first proposed as a low-cost alternative to LISA in the 1990s. In China, a GW mission in Earth orbit called TIANQIN[82] has been proposed and under study.

Table 3 lists the orbit configuration, arm length, orbit period and S/C number of various GW space mission proposals.

Table 3. A Compilation of GW Mission Proposals

| Mission Concept | S/C Configuration | Arm length | Orbit Period | S/C # |
|---|---|---|---|---|
| *Solar-Orbit GW Mission Proposals* | | | | |
| LISA[65] | Earth-like solar orbits with 20° lag | 5 Gm | 1 year | 3 |
| eLISA[64] | Earth-like solar orbits with 10° lag | 1 Gm | 1 year | 3 |
| ASTROD-GW[68] | Near Sun-Earth L3, L4, L5 points | 260 Gm | 1 year | 3 |
| Big Bang Observer[73] | Earth-like solar orbits | 0.05 Gm | 1 year | 12 |
| DECIGO[72] | Earth-like solar orbits | 0.001 Gm | 1 year | 12 |
| ALIA[74] | Earth-like solar orbits | 0.5 Gm | 1 year | 3 |
| ALIA-descope[75] | Earth-like solar orbits | 3 Gm | 1 year | 3 |
| Super-ASTROD[71] | Near Sun-Jupiter L3, L4, L5 points (3 S/C), Jupiter-like solar orbit(s)(1-2 S/C) | 1300 Gm | 11 year | 4 or 5 |
| *Earth-Orbit GW Mission Proposals* | | | | |
| OMEGA[81] | 0.6 Gm height orbit | 1 Gm | 53.2 days | 6 |
| gLISA/GEOGRAWI[76-78] | Geostationary orbit | 0.073 Gm | 24 hours | 3 |
| GADFLI[79] | Geostationary orbit | 0.073 Gm | 24 hours | 3 |
| TIANQIN[82] | 0.057 Gm height orbit | 0.11 Gm | 44 hours | 3 |
| ASTROD-EM[69,70] | Near Earth-Moon L3, L4, L5 points | 0.66 Gm | 27.3 days | 3 |
| LAGRANGE[80] | Near Earth-Moon L3, L4, L5 points | 0.66 Gm | 27.3 days | 3 |

Typical frequency sensitivity spectrum of strain for space GW detection consists of 3 regions (Fig. 3), the acceleration noise region, the shot noise (flat for current space detector projects like LISA in strain psd) region, if any, and the antenna response region. The lower frequency region for the detector sensitivity is dominated by vibration, acceleration noise or gravity-gradient noise. The higher frequency part of the detector sensitivity is restricted by antenna response (or storage time). In a power-limited design, sometimes there is a middle flat region in which the sensitivity is limited by the photon shot noise.[10,65,84]

The shot noise sensitivity limit in the strain for GW detection is inversely proportional to $P^{1/2}l$ with $P$ the received power and $l$ the distance. Since $P$ is inversely proportional to $l^2$ and $P^{1/2}l$ is constant, this sensitivity limit is independent of the distance. For 1-2 W emitting power, the limit is around $10^{-21}$ Hz$^{-1/2}$. As noted in the LISA study,[65] making the arms longer shifts the time-integrated sensitivity curve to lower frequencies while leaving the bottom of the curve at the same level. Hence, ASTROD-GW with longer arm length has better sensitivity at lower frequency. e-LISA and GW interferometers in Earth orbit have shorter arms and therefore have better sensitivities at higher frequency.

In Fig.'s 2-4, we plot sensitivity curves for LISA, e-LISA and ASTROD-GW for the low-frequency GW band. In the Mock LISA Data Challenge (MLDC) program, the consensus goal for the LISA instrumental noise density amplitude $^{(MLDC)}S_{Ln}^{1/2}(f)$ is

$$^{(MLDC)}S_{Ln}^{1/2}(f) = (1/L_L) \times \{[(1 + 0.5 \ (f/f_L)^2\ )] \times S_{Lp} + [1 + (10^{-4}\ \text{Hz}\ /f)^2]\ (4S_a/(2\pi f)^4)\}^{1/2}\ \text{Hz}^{-1/2},\ (68)$$

where $L_L = 5 \times 10^9$ m is the LISA arm length, $f_L = c\ /\ (2\pi L_L)$ is the LISA arm transfer frequency, $S_{Lp} = 4 \times 10^{-22}$ m$^2$ Hz$^{-1}$ is the LISA (white) position noise level due to photon shot noise, and $S_a = 9 \times 10^{-30}$ m$^2$ s$^{-4}$ Hz$^{-1}$ is the LISA white acceleration noise (power) level.[85] Note that (68) contains the "reddening" factor $[1 + (10^{-4}\ / f)^2]$ in the acceleration noise term.

If we drop the "reddening factor", the enhanced LISA instrumental noise density amplitude $^{(Enhanced)}S_{Ln}^{1/2}(f)$ becomes

$$^{(Enhanced)}S_{Ln}^{1/2}(f) = (1/L_L) \times \{[(1 + 0.5\ (f/f_L)^2)] \times S_{Lp} + [4S_a/(2\pi f)^4]\}^{1/2}\ Hz^{-1/2}. \qquad (69)$$

The eLISA arm length $L_{eL}$ is 5 times shorter. Its instrumental noise density amplitude $^{(MDLC)}S_{eLn}^{1/2}(f)$ is

$$^{(MDLC)}S_{eLn}^{1/2}(f) = (1/L_{eL}) \times \{[(1 + 0.5\ (f/f_{eL})^2)] \times S_{eLp} + [1+(10^{-4}\ Hz\ /f)^2](4S_a/(2\pi f)^4)\}^{1/2}\ Hz^{-1/2}, \quad (70)$$

where $L_{eL} = 10^9$ m is the eLISA arm length, $f_{eL} = c\ /\ (2\pi L_{eL})$ is the $e$LISA arm transfer frequency, $S_{eLp} = 1 \times 10^{-22}$ m$^2$ Hz$^{-1}$ is the eLISA (white) position noise level due to photon shot noise assuming that the telescope diameter is 25 cm (compared with 40 cm for that of LISA) and that the laser power is the same as LISA. The corresponding enhanced eLISA instrumental noise density amplitude $^{(Enhanced)}S_{eLn}^{1/2}(f)$ is

$$^{(MDLC)}S_{eLn}^{1/2}(f) = (1/L_{eL}) \times \{[(1 + 0.5\ (f/f_{eL})^2)] \times S_{eLp} + (4S_a/(2\pi f)^4)\}^{1/2}\ Hz^{-1/2}. \qquad (71)$$

For ASTROD-GW, our goal on the instrumental strain noise density amplitude is

$$S_{An}^{1/2}(f) = (1/L_A) \times \{[(1 + 0.5\ (f/f_A)^2)] \times S_{Ap} + [4S_a/(2\pi f)^4]\}^{1/2}\ Hz^{-1/2}, \qquad (72)$$

over the frequency range of 100 nHz $< f <$ 1 Hz. Here $L_A = 260 \times 10^9$ m is the ASTROD-GW arm length, $f_A = c\ /\ (2\pi L_A)$ is the ASTROD-GW arm transfer frequency, $S_a = 9 \times 10^{-30}$ m$^2$ s$^{-4}$ Hz$^{-1}$ is the white acceleration noise level (the same as that for LISA), and $S_{Ap} = 10816 \times 10^{-22}$ m$^2$ Hz$^{-1}$ is the (white) position noise level due to laser shot noise which is 2704 (=52$^2$) times that for LISA.[13,14,16,68,86] The corresponding noise curve for the ASTROD-GW instrumental noise density amplitude $^{(MDLC)}S_{An}^{1/2}(f)$ with the same "reddening" factor as specified in MLDC program is

$$S_{An}^{1/2}(f) = (1/L_A) \times \{[(1 + 0.5\ (f/f_A)^2)] \times S_{Ap} + [1 + (10^{-4}/f)^2]\ (4S_a/(2\pi f)^4)\}^{1/2}\ Hz^{-1/2}, \quad (73)$$

over the frequency range of 100 nHz $< f <$ 1 Hz. The sensitivity curves from the six formulas (68) to (73) are shown in Fig. 3. The corresponding sensitivity curves in terms of $h_c(f)$ and $\Omega_{gw}(f)$ are shown in Fig. 2 and Fig. 4 respectively.

The LISA Pathfinder Mission has been shipped to Kourou and is scheduled for launch from Kourou Spaceport on Arianespace Flight VV06 on 2nd December 2015. It is a technology demonstration mission. Its success will pave the road for future space GW missions.

The sensitivity curve of a single DECIGO interferometer as shown in Fig. 3 is from [87]. BBO has a similar single-interferometer sensitivity curve. One-sigma, power-law integrated sensitivity curve for BBO (BBO-corr) as shown in Fig. 3 is obtained by Thrane and Romano [88]. That of DECIGO is similar. We also put in the plot their LISA autocorrelation measurement sensitivity curve (LISA-corr) in a single detector assuming perfect subtraction of instrumental noise and/or any unwanted astrophysical foreground.[88] The minimum autocorrelation sensitivity using the same method for ASTROD-GW is also estimated and plotted in Fig. 3; this would also be the level that 6 S/C ASTROD-GW[68] (6 S/C ASTROD-GW-corr) could reach. For comparison, the one-sigma, power-law integrated sensitivity curve for the adLIGO H1L1 (adLIGO-corr) from ref. [88] is also plotted in Fig. 3. All of the corresponding curves are plotted in Fig. 2 and Fig. 4.

The development in atom interferometry is fast and promising. It already contributes to precision measurement and fundamental physics. A proposal using atom interferometry to detect GWs has been raised at Stanford University as an alternate method to LISA on the LISA bandwidth.[89,90] Issues have arisen on its realization of LISA sensitivity.[91,92] In Observatoire de Paris, SYRTE has started the first stage its project -- MIGA (Matter-wave laser Interferometric Gravitation Antenna)[93] of building a 300-meter long optical cavity to interrogate atom interferometers at the underground laboratory LSBB in Rustrel. In the second stage of the project (2018-2023), MIGA will be dedicated to science runs and data analyses in order to probe the spatio-temporal structure of the local field of the LSBB region. In the meantime, MIGA will assess future potential applications of atom interferometry to GW detection in the middle frequency band (0.1-10 Hz).

### 3.6. Very low frequency band (300 pHz – 100 nHz)

When GWs are propagating across the line of sight of pulsar observations, the pulse arrival times are affected. This effect can be used to observe the GWs. For isotropic stochastic GW background, Hellings and Downs derived a formula on the correlation in the timing residuals as a function of pairs of pulsars and used it to constrain the energy density in GWs of frequency between 4-10 nHz to be less than $1.4 \times 10^{-4}$ of the cosmic critical density in 1983.[23] In 1996 and 2002, the upper limits from pulsar timing observations on a GW background derived by McHugh $et\ al.$[94] and by Lommen[95] are $\Omega_{gw} \lesssim 10^{-7}$ in the frequency range 4-40 nHz, and $\Omega_{gw} \leq 4 \times 10^{-9}$ at $6 \times 10^{-8}$ Hz respectively.

Now there are 4 major pulsar timing arrays (PTAs): the European PTA (EPTA),[96] the NANOGrav,[97] the Parks PTA (PPTA)[98] and the International (EPTA, NANOGrav and PPTA combined) PTA (IPTA).[99] For recent reviews on pulsar timing and pulsar timing arrays for GW detection, please see [100, 101]. These 4 PTAs have improved greatly on the sensitivity for GW detection recently.[102-104] Upper limits on the isotropic stochastic background from EPTA, PPTA and NANOGrav are listed in Table 4. These limits assumes that the GW background has the following frequency dependence with $\alpha = -(2/3)$;

$$h_c(f) = A_{yr}\ [f/(1\ yr^{-1})]^{\alpha}. \tag{74}$$

The most stringent limit is from Shannon $et\ al.$[103] using observations of millisecond pulsars from the Parks telescope to constrain $A_{yr}$ to less than $1.0 \times 10^{-15}$ with 95 % confidence. This limit already excludes present and most recent model predictions with 91-99.7 % probability.[103] The three experiments form a robust upper limit of $1 \times 10^{-15}$ on $A_{yr}$ at 95 % confidence level ruling out most models of supermassive black hole formation. The limit is shown as constraint on the Supermassive Black Hole Binary GW Background (SBHB-GWB) in Fig. 2; the corresponding constraints are also shown in Fig. 3 and Fig. 4. Since more energy of GWs might be emitted with higher frequency in the hierarchy of supermassive black hole formation, we extrapolate this constraint linearly with dotted line to $1 \times 10^{-5}$ Hz with some confidence. Constraints with other $\alpha$ values have similar order of magnitudes.

Table 4. Upper limits on the isotropic stochastic background from 3 pulsar timing arrays.

| | No. of pulsars included | No. of years observed | Observation radio band [MHz] | Constraint on characteristic strain $h_c(f)$ [= $A_{yr}$ [$f/(1\,yr^{-1})$]$^{-(2/3)}$, ($f = 10^{-9}$-$10^{-7}$ Hz)] |
|---|---|---|---|---|
| EPTA[102] | 6 | 18 | 120-3000 | $A_{yr} < 3 \times 10^{-15}$ |
| PPTA[103] | 4 | 11 | 3100 | $A_{yr} < 1 \times 10^{-15}$ |
| NANOGrav[104] | 27 | 9 | 327-2100 | $A_{yr} < 1.5 \times 10^{-15}$ |

To have an outlook of sensitivity of PTAs for the next hundred years, we adopt and extrapolate the estimates of Moore, Taylor and Gair.[105] The sensitivity for a monochromatic GW of a PTA is mainly dependent on the timing accuracies including timing residuals after modelling (rms deviations in timing residuals). The bandwidth depends on the sampling frequencies, i.e. cadences and the duration of the data span. For observations every $\Delta t$ of time and an observation span of $T$ the bandwidth $f$ is [$1/T$, $1/\Delta t$]. The frequency dependence of the sensitivity in $h_c(f)$ is linear in $f$:

$$h_c(f) = B_{yr}\,(f\,/\,yr^{-1}), \quad (1/T) < f < (1/\Delta t). \tag{75}$$

We assume $B_{yr}$ is proportional to the timing accuracy, inversely proportional to the observation time span and inversely proportional to the number of pulsars in the PTA. In Moore, Taylor and Gair [105], a canonical PTA (MTG canonical PTA) with 36 pulsars randomly distributed on the sky, observed every two weeks with a precision of 100 ns over 5 years was assumed; this canonical PTA has a sensitivity (75) with $B_{yr} = 4 \times 10^{-16}$ and is roughly equivalent to OPEN 1 mock dataset in the IPTA data challenge.[106] In Table 5, we compile the projected sensitivities for IPTA, FAST[107] and SKA[108] for an observation span of 20 years, 50 years and 100 years respectively. To obtain a fiducial sensitivity of IPTA, we take the MTG canonical PTA [103], but extend the observation time span to 20 yrs. The sensitivity is $1 \times 10^{-16}$ at $f = yr^{-1} = 3.17 \times 10^{-8}$. With the advent of new and more sensitive observing facilities, PTA sensitivity will be improved. The Five-Hundred-Meter Aperture Spherical Radio Telescope (FAST)[107] is under construction in Guei-Zhou, China. Since the 305 m radio telescope of Arecibo Observatory has been working for 52 years, we expect that FAST will work for more than 50 years also. In obtaining a fiducial sensitivity, we assume the FAST PTA to observe 50 pulsars with 50 ns timing accuracy for a 50 yr time span. The Square Kilometre Array (SKA)[108] in South Africa and Australia will certainly improve on existing limits and we assume pulsar timing measurements every 2 weeks for 100 pulsars with 20 ns timing accuracies for 100 yrs. Table 5 lists the basic assumptions for IPTA, FAST and SKA and their projected sensitivities in $B_{yr}$ on the characteristic strain.

Table 5. Sensitivities of IPTA, FAST and SKA to monochromatic GWs.

| | No. of pulsars | No. of years of observation | Timing accuracy (ns) | Sensitivity in characteristic strain $h_c(f)$ [= $B_{yr}\,(f\,/\,yr^{-1})$] for monochromatic GWs |
|---|---|---|---|---|
| IPTA[106] | 36 | 20 | 100 | $B_{yr} = 1 \times 10^{-16}$ |
| FAST[107] | 50 | 50 | 50 | $B_{yr} = 1.5 \times 10^{-17}$ |
| SKA[108] | 100 | 100 | 20 | $B_{yr} = 1.5 \times 10^{-18}$ |

The sensitivity curves of IPTA, FAST and SKA:

$$h_c(f) = 1 \times 10^{-16} \, (f \,/\, \mathrm{yr}^{-1}), \quad 1.58 \times 10^{-9} \, \mathrm{Hz} < f < 8.27 \times 10^{-7} \, \mathrm{Hz}, \quad \text{for IPTA-20,}$$

$$h_c(f) = 1.5 \times 10^{-17} \, (f \,/\, \mathrm{yr}^{-1}), \quad 6.34 \times 10^{-10} \, \mathrm{Hz} < f < 8.27 \times 10^{-7} \, \mathrm{Hz}, \quad \text{for FAST-50,}$$

$$h_c(f) = 1.5 \times 10^{-18} \, (f \,/\, \mathrm{yr}^{-1}), \quad 3.17 \times 10^{-10} \, \mathrm{Hz} < f < 8.27 \times 10^{-7} \, \mathrm{Hz \, Hz}, \quad \text{for SKA-100,}$$

are plotted on Fig. 2. The corresponding sensitivity curves in terms of $[S_h(f)]^{1/2}$ and $\Omega_g(f)$ are plotted in Fig. 3 and Fig. 4 respectively. We notice that the SKA sensitivity for monochromatic GWs reaches $10^{-22}$ in $\Omega_g(f)$ at frequency around $3.17 \times 10^{-10}$ Hz. The acronyms for these curves are IPTA-20, FAST-50 and SKA-100.

As to the single source GW limits. The bounds from PPTA[109] and EPTA[110] are in the order of $10^{-14}$ for $h_c$ in the frequency range $5 \times 10^{-9}$ to $2 \times 10^{-7}$. They are drawn on Fig. 2 with the corresponding curves on Fig.3 and Fig. 4. A 24-hr global campaign for GW from J1713+0747 gives upper limits in the frequency range $10^{-5}$-$10^{-3}$ Hz; the solid line shows the upper limit in random direction while the dotted line show the upper limit in the direction of pulsars.[111]

### 3.7. Ultralow frequency band (10 fHz – 300 pHz)

GWs with periods longer than the time span of observations produce a simple pattern of apparent proper motions over the sky.[112] Therefore, precise measurement of proper motion of quasars would be a method to detect ultra-low frequency (10 fHz – 300 pHz) gravitational waves. Gwinn *et al.*[113] used this method to constrain the normalized spectral energy density of stochastic GWs with frequencies less than $2 \times 10^{-9}$ Hz and greater than $3 \times 10^{-18}$ Hz (including frequencies in the ultra-low frequency band) to less than 0.11 $h^{-2}$ (95 % confidence) times the critical closure density of our Universe. In Fig. 4, we use the Planck 2015 value [25] of Hubble constant $H_0 = (67.8 \pm 0.9)$ km s$^{-1}$Mpc$^{-1}$ to set $h = 0.678$ in their original plot and obtain a bound of 0.24 in terms of the critical density (the bound is labelled QA [Quasar Astrometry] in Fig. 4). Long baseline optical interferometer with sub-micro-arcsecond and nano-arcsecond (nas) astrometry is technologically feasible.[114] With this kind of interferometer implemented, precision astrometry of quasar proper motions may possibly be improved by 4 orders of magnitude and reach nas yr$^{-1}$. In terms of energy, the precision of determining/constraining $\Omega_{gw}(f)$ could reach a sensitivity of $2.4 \times 10^{-9}$ or better (Fig. 4; the curve is labelled QAG [Quasar Astrometry Goal]).

Using (60) or Table 2, we have the bound on characteristic strain $h_c(f)$:

$$h_c(f) < 4.2 \times 10^{-19} \, (\mathrm{Hz}/f), \quad \text{for } 3 \times 10^{-18} \, \mathrm{Hz} < f < 2 \times 10^{-9} \, \mathrm{Hz}. \tag{76}$$

When the angle resolution is improved by 4 orders of magnitude, the sensitivity reaches

$$h_c(f) = 4.2 \times 10^{-23} \, (\mathrm{Hz}/f), \quad \text{for } 3 \times 10^{-18} \, \mathrm{Hz} < f < 2 \times 10^{-9} \, \mathrm{Hz}. \tag{77}$$

Both the bound (76) and the curve (77) are plotted on Fig. 2. They are labelled QA and QAG. Using (60), we also convert $\Omega_{gw}(f)$ to $[S_h(f)]^{1/2}$ and plot the results on Fig. 3.

### 3.8. Extremely low (Hubble) frequency band (1 aHz – 10 fHz)

The successful prediction of nucleosynthesis of primordial abundances of $^3$He, $^4$He, $^7$Li and deuterium put a constraint of integrated tensor perturbations $\int d(\log f)\Omega_{gw}(f)$ of $10^{-5}$.[115] This is plotted on Fig. 4 as the $\Omega_{gw}(f) = 10^{-5}$ line. Cosmic microwave background (CMB) experiments are most sensitive to the extremely low (Hubble) frequency band (1 aHz − 10 fHz). First, a strong GW background at extremely low frequency produces stochastic redshift of CMB (Sachs-Wolfe effect).[116,117] The COBE observation gives CMB S-W redshift fluctuation bound which was plotted on Fig's 2-4 as CMB S-W fluct. The COBE microwave-background quadrupole anisotropy measurement[118,119] gives a limit $\Omega_{gw}$ (1 aHz) ~ $10^{-9}$ on the extremely-low-frequency GW background.[120,121] WMAP[122,123,124] improves on the COBE constraints; the constraint on $\Omega_{gw}$ for the higher frequency end of this band is better than $10^{-14}$. Planck Surveyor space mission has recently probed anisotropies with $l$ up to 2000 and with higher sensitivity. Ground and balloon experiments probe smaller-angle anisotropies and, hence, higher-frequency background. ACTpol has probed anisotropies with $l$ from 225 up to 8725.[125] These CMB experiments probe the 1 aHz − 10 fHz extremely low (Hubble) frequency band GWs. In inflationary cosmology these GWs give the tensor mode density and temperature perturbations (imprints) on CMB.

Inflation postulates a rapid accelerated expansion which set the initial moments of the Big Bang Cosmology.[126-131] Expansion drives the universe towards a homogeneous and spatially flat geometry that accurately describes the average state of the universe. The quantum fluctuations in this era grow into the galaxies, clusters of galaxies and temperature anisotropies of the cosmic microwave background.[132-137] Modern inflation has been originated from efforts of unification, but its mechanism still remains unclear. The quantum fluctuations in the space-time geometry in the inflationary era generate GWs which would have imprinted tensor perturbations on the CMB anisotropy. There is no confirmed discovery of these tensor perturbations yet ([138], [139] and references therein). The analysis of *Planck*, SPT, and ACT temperature data together with WMAP polarization did not discover these tensor perturbations and showed that the scalar index is $n_s = 0.959 \pm 0.007$ and the tensor-to-scalar perturbation ratio $r$ is less than 0.11 (95% CL; no running).[140] The pivot scale of this constraint is 0.002 $Mpc^{-1}$, corresponding to GW frequency $f$ ~ $1.5 \times 10^{-18}$ Hz at present. From Ref.'s [141,117], this constraint corresponds to $\Omega_{gw} < 10^{-15}$ and $h_c < 2.34 \times 10^{-9}$; it is plotted on Fig.'s 2-4 with label *Planck*. In March 2014, the announcement of BICEP2 of the detection of B-mode polarization excess and their interpretation of this excess as imprint from primordial tensor waves immediately attracted the imagination of the scientific community and the public.[142] The September 2014 announcement of dust measurement including the BICEP2 observation region from the Planck team convinced the physics community that the excess is consistent with dust emission [143]. The new *Keck Array* data [138] confirmed the BICEP2 B-mode polarization excess. The combined analysis of BICEP2/*Keck Array* and *Planck* Collaboration [138] convincingly showed that this excess is consistent with the *Planck* dust measurement and that the tensor-to-scalar perturbation ratio $r$ is constrained to less than 0.12 (95% CL; no running). The pivot scale of this constraint is 0.05 $Mpc^{-1}$, corresponding to GW frequency $f$ ~ $3.8 \times 10^{-17}$ Hz at present. From [141,117], this constraint corresponds $\Omega_{gw} < 1.1 \times 10^{-15}$ and $h_c < 5.57 \times 10^{-8}$; it is plotted on Fig.'s 2-4 with label BICEP2/*Keck*.

The sources for B-mode polarization in CMB could come from (i) GW imprints on CMB; (ii) Gravitational lensing during the CMB propagation; and (iii) Pseudo-scalar-photon interaction during the CMB propagation. In the BICEP2 analysis [142], gravitational lensing effect is subtracted. Einstein equivalence principle dictates that the propagation of electromagnetic waves (photons) observes Maxwell equations locally and there is no rotation of polarization plane during propagation (i.e. no comic polarization rotation [CPR]). However, this is exactly a soft spot in the empirical foundation of EEP [144]. For a survey of constraints on CPR from astrophysical and cosmological observations, see [145]. Basically, both the mean CPR and the CPR fluctuation magnitude are constrained to a couple of degrees. For example, from the ACTpol CMB polarization data fitting [146], the mean CPR angle $\alpha$ is constrained from the EB correlation power spectra to be less than about 1° and the fluctuation (rms) is constrained from the BB correlation power spectra to $<\delta\alpha^2>^{1/2} < 1.68°$. Including CPR effect together with the Planck dust measurement in a joint fitting of ACTpol, BICEP2, and POLARBEAR gives the values of the mean squares of the CPR fluctuation $<\delta\alpha^2> = 41 \pm 522$ [mrad²] and the tensor-to-scalar ratio r = −0.012 ± 0.109; this in turn gives a 1 σ bound on the rms of the CPR magnitude $<\delta\alpha^2>^{1/2} < 23.7$ mrad (1°.36) and that of r < 0.097. This result not only gives the best constraint on the CPR fluctuation magnitude, but also is consistent with the Planck and the joint BICEP2/*Keck Array*-Planck bound.

The ongoing situation as said in view given by Halverson [139] is "The competition is fierce, with at least six funded ground-based experiments underway (including the third version of BICEP), several balloon-borne experiments, and a number of proposed space missions. Finally, thanks to the new data, galactic foreground contaminants—and strategies for removing them—are now better understood." The present consensus is that when the present ground-based and balloon-borne experiments are performed, the accuracy in determine *r* will have one order of magnitude improvement to 0.01; when the proposed space missions are flown and completed, the accuracy will have another order of magnitude improvement to 0.001. This means that the sensitivity in the $\Omega_{\text{gw}}$-*f* plot will be improved to $10^{-17}$.

## 4. Sources of gravitational waves

In this section, we discuss sources of GWs concisely while refer to various references for more complete treatment.

### 4.1. GWs from compact binaries

Binary neutron stars coalesce by losing kinetic energy of orbital motion due to the emission of gravitational wave. When the orbital radius is much larger than the radius of stars, the radiation of gravitational wave is described by the quadrupole approximation reviewed in Sec. 2 until merging starts where two stars are deformed by each other's tidal forces.[147] The wave signal chirps according to the advancement of time and the frequency ranges from the low frequency orbital motion period to high frequency merger characteristic frequency ($\sim 1$ kHz). Since the amplitude of the signal increases till the merger (inspiral phase), the signal of this inspiral phase is the most probable target of all ground based laser interferometers with km-scale baseline length.

There are several neutron star binaries in our Galaxy (in the case of J1906+0746 the companion star may be a white dwarf). In Table 3, all that may coalesce due to the emission of gravitational wave within the age of the universe are listed. After the merger, coalesced neutron stars form a black hole and it oscillates due to dynamical energy of the coalescence just after the merger, which is known as quasi-normal mode oscillation of black hole. Since its typical frequency is several kHz for black hole with a few $M_\odot$, the oscillation (ring down) is the gravitational wave target of detectors that have sensitivity at high frequencies such as GEO-HF detector or cryogenic mechanical detectors.

Table 6. Neutron star binaries in our galaxy that may coalesce within the age of the Universe (Companion star of J1906+0746 may be a white dwarf). $P_s$ is the pulse period, $P_b$ the orbital period of the binary system, $e$ the eccentricity of the orbit, and $\tau_{life}$ the life time of the binary system.

| | $P_s$ (ms) | $P_b$ (hr) | $e$ | $\tau_{life}$ (Gyr) |
|---|---|---|---|---|
| B1913+16[148] | 59.03 | 7.75 | 0.62 | 0.37 |
| B1534+12[149] | 37.40 | 10.10 | 0.27 | 2.93 |
| J0737-3039A[150] | 22.70 | 2.45 | 0.088 | 0.23 |
| J1756-2251[151] | 28.46 | 7.67 | 0.18 | 2.03 |
| J1906+0746[152] | 144.14 | 3.98 | 0.085 | 0.082 |
| J2127+11C[153] | 32.76 | 8.04 | 0.68 | 0.32 |

The coalescence rate of binary neutron stars is estimated by knowing both the distribution of binaries and the life time of the binary systems. Due to the small number statistics and to the uncertainty biases of pulsar observation (e.g. dissipation of electromagnetic waves in our Galaxy, beaming angle, faint pulsars, etc.), the estimated event rate ranges more than 2 orders of magnitude, where typical value is once per 100 thousand years in such matured galaxy as ours.[154] Since the population of such matured galaxy is roughly 0.01 per cubic Mpc, at least one event per year can be detected if the sensitivity to catch events occurring at 130 Mpc is achieved by ground based detector. Advanced LIGO has initiated observation run 1 (O1) starting September 18, 2015 with sensitivity reaching 70 Mpc for the coalescence of nominal binary neutron stars.

In the coalescence of binary black holes, the frequency of the chirping signal shifts down to lower frequencies. If their initial mass ranges around 10 $M_\odot$, the merger may occur at around 200 Hz. The signal is in the most sensitive frequencies in the second generation ground based interferometers.

Dominick estimated that the population and the coalescence rate of binary black holes is smaller than that of binary neutron stars.[155] However, a theoretical study shows that merger rate of black holes based on ejections from globular clusters is larger than that of neutron star binaries.[156] This is still an issue of different opinions. Moreover, since the amplitude of gravitational waves from the coalescence of black holes is larger, possible detection rate will be larger if the detector has sensitivity at lower frequencies (< ~10 Hz), which will be realized by the third-generation detectors.

We plot the source strengths of compact binary inspirals, pulsars, resolvable galactic binaries and unresolvable galactic binaries [confusion background][3,157] in Fig.'s 2-4 by adopting those of Moore, Cole and Berry [35].

### 4.2. GWs from supernovae

Massive stars heavier than 8 $M_\odot$ collapse due to gravity after burning out and a neutron star may be born. This collapse produces burst gravitational wave. Taking the second derivative of the quadrupole moment of the star and using (35), the maximum amplitude $h_{max}$ is estimated to be $\kappa MR^2(2\pi f)^2$, where $\kappa$ [of the order of $G_N/(c^6 r)$ times non-sphericity of the explosion] is a calculable numerical factor, $M$ is the mass of the initial neutron star born just after the collapse, $R$ is the radius of the star, and $r$ is the observation distance to the star. If the collapse occurs in the center of our galaxy in a favorable condition, the burst wave signal may be detected by resonant antennas. And also it is a target source of gravitational wave of ground based interferometric detectors.[158]

The gravitational wave form information is useful to enhance the signal-to-noise ratio of the detector. Since stellar core collapse is a complex physical phenomena that involves general relativity, hydrodynamics, and neutrino transport with thermonuclear kinetics in short time duration, it is not easy to conduct a full numerical simulation to obtain the gravitational wave form. In 2002 Dimmelmeier, Font, and Müller[159] first performed axisymmetric hydrodynamic simulations of rotational core collapse and its associated GW emission in 26 general-relativistic and Newtonian models. The total energy of GWs emitted is only about $10^{-7}$-$10^{-8}$ M$_\odot c^2$. Recent development in three dimensional numerical simulation which requires longer computing time shows that strong burst GW of total luminosity of 0.01 M$_\odot c^2$ can be produced by an initially non-rotating star due to standing accretion shock instability (SASI).[160,161] For a review on this subject, see [162]. These may be the plausible candidates for the second-generation ground based detectors. The event rate of supernova explosions in our Galaxy is estimated as once per $40 \pm 10$ yr.[163] According to Abadie *et al.*,[164] supernova explosion with GW energy 0.056 M$_\odot c^2$ could be detected at 16 Mpc with LIGO-Virgo achieved sensitivity; hence supernova explosion with GW energy ~0.01 M$_\odot c^2$ should be detected up to 6.8 Mpc with the LIGO-Virgo achieved sensitivity; if supernova explosion is always accompanied with the emission of GW energy of ~0.01 M$_\odot c^2$, the detection rate on the Earth would be 0.04/yr assuming uniform distribution of such galaxies as ours to be 0.01 Mpc$^{-3}$. This rate would nominally be improved to 1.7 yr$^{-1}$ by adLIGO at present sensitivity (3.5 fold improvement compared with [164]). However, since there is a large uncertainty in the distribution of GW energy strength in the supernova explosion, we just adopting the strength as given in Moore, Cole and Berry [35] for plotting in Fig.'s 2-4.

### 4.3. GWs from massive black holes and their coevolution with galaxies

Observational evidences indicate that massive black holes (MBHs) residing in most local galaxies. Relations have been discovered between the MBH mass and the mass of host galaxy bulge, and between the MBH mass and the velocity-dispersion. These relations indicate that the central MBHs are linked to the evolution of galactic structure. Newly fueled quasar may come from the gas-rich major merger of two massive galaxies. Recent astrophysical evidences linked together these major galaxy mergers and the growth of supermassive black holes in quasars.[165,166] Distant quasar observations indicate that MBH of billions of solar masses already existed less than a billion years after the Big Bang. At present, there are different theoretical proposals for scenarios of the initial conditions and formations of black holes. These scenarios include BH seeds from inflationary Universe and/or from the collapse of Population III stars, different accretion models and binary formation rates. All of these models generate MBH merging scenarios in galaxy co-evolution with GW radiations. Measurement of amplitude and spectrum of these GWs will tell us the cosmic history of MBH formation.

The standard theory of MBH formation is the merger-tree theory with various Massive Black Hole Binary (MBHB) inspirals acting. The GWs from these MBHB inspirals can be detected and explored to cosmological distances using space GW detectors. Although there are different merger-tree models and models with BH seeds, they all give significant detection rates for space GW detectors and PTAs.[167-169,100] Gravitational wave (GW) observation in the 300 pHz – 0.1 Hz frequency band will be a major observation tool to study the co-evolution of galaxy with BHs. This frequency band covers the low frequency band (100 nHz - 100 mHz) and very low frequency band (300 pHz-100 nHz) GWs and is in the detection range of PTAs, eLISA and ASTROD-GW. PTAs are most sensitive in the frequency range 300 pHz -100 nHz, eLISA space GW detector is most sensitive in the frequency range 2 mHz – 0.1 Hz, while ASTROD-GW is most sensitive in the frequency range 500 nHz - 2 mHz (Fig.'s 2-4).

PTAs have been collecting data for decades for detection of stochastic GW background from MBH binary mergers. They already constrain $A_{yr}$ in (72) to less than $1.0 \times 10^{-15}$ with 95 % confidence.[102-104] This limit excludes present and most recent model predictions of supermassive black hole formation with 91-99.7 % probability.[103] This means the detection could be anytime near. Since we know that SMBHs are already formed, it also means that the backgrounds in the higher frequency/shorter wavelength band are higher than original predicted. For most models there is a knee around $f \sim 100$ nHz, now we straighten the knee and extend (72) to $f \sim 10$ µHz with dashed line in Fig. 2. Below this $1.0 \times 10^{-15}$ limit, we plot pink colored region to show possible background source region. Corresponding line and colored region are also shown in Fig. 3 and Fig. 4.

eLISA and ASTROD-GW will be able to directly observe how massive black holes form, grow, and interact over the entire history of galaxy formation. ASTROD-GW will detect stochastic GW background from MBH binary mergers in the frequency range 500 nHz to 100 µHz. These observations are significant and important to the study of co-evolution of galaxies with MBHs. The expected rate of MBHB sources is 10 $yr^{-1}$ to 100 $yr^{-1}$ for eLISA and 10 $yr^{-1}$ to 1000 $yr^{-1}$ for LISA.[64] For ASTROD-GW, similar number of sources as that of LISA is expected with better angular resolution.[68] For a more detailed account, see [170].

At present, there are different theoretical scenarios for the initial conditions and formations of black holes, e.g., primordial massive BH clouds as seeds, direct formation of supermassive black hole via multi-scale gas inflows in galaxy mergers, direct collapse into a supermassive black hole from mergers between massive protogalaxies with no need to suppress cooling and star formation, etc. The mass range and maximum mass of Population III stars is also a relevant issue for seed BHs. With the PPTA constraint, there should be more backgrounds in the µHz region. ASTROD-GW with good sensitivity in the µHz band will contribute to detect or constrain GW background to distinguish various scenarios for finding the history of BH and galaxy co-evolution.

With the detection of MBH merger events and background, the properties and distribution of MBHs could be deduced and underlying population models could be tested. Sesana *et al*.[171] consider and compare ten specific models of massive black hole formation. These models are chosen to probe four important and largely unconstrained aspects of input physics used in the structure formation simulations, i.e., seed formation, metallicity feedback, accretion efficiency and accretion geometry. With Bayesian analyses to recover posterior probability distribution, they show that LISA has enormous potential to probe the underlying physics of structure formation. With better sensitivity in the frequency range 100 nHz - 1 mHz, ASTROD-GW will be able to probe the underlying physics of structure formation further. With the detection of the GW background of the MBH mergers, PTAs and ASTROD-GW will add to our understanding of the structure formation.

We plot the source strengths of massive binaries in Fig.'s 2-4 adopting those of Moore, Cole and Berry [35].

### 4.4. GWs from extreme mass ratio inspirals (EMRIs)

EMRIs are GW sources for space GW detectors. The eLISA sensitive range for central MBH masses is $10^4$-$10^7$ M$_\odot$. The expected number of eLISA detections over two years is 10 to 20;[64] for LISA, a few tens;[64] for ASTROD-GW, similar or more with sensitivity toward larger central BH's and with better angular resolution.[68] For a more detailed account, we refer to Ref.'s [170]. We plot the source strengths of EMRIs in Fig.'s 2-4 adopting those in Moore, Cole and Berry [35].

### 4.5. Primordial/inflationary/relic GWs

Relic GWs from inflationary or non-inflationary period are commonly called primordial GWs. Relative to primordial GWs, all the GW sources we have discussed are foregrounds. Assuming the primordial GW spectrum is flat in the $\Omega_{gw}(f)$ vs $f$ diagram, i.e. the tensor index $n_t$ is zero, we draw an upper bound of inflationary spectrum to saturate the constraints given in Sec. 3.8; it is the flat line (the tensor index $n_t$ is zero) about $10^{-15}$ level in the $\Omega_{gw}(f)$ vs $f$ diagram (Fig. 4) with the very high frequency part dropping steeply above $10^{10}$ Hz. For comparison, the black dotted curve shows the corresponding $\Omega_{gw}(f)$ for a 0.9 K blackbody radiation. If the GW perturbations had been in equilibrium with the matter fields, it is an expected GW background. We refer the readers to the recent review by Sato and Yokoyama on "Inflationary cosmology: First 30+ years"[172] for a detailed account of the inflationary scenario.

As expected in Sec. 3.8, the present consensus on the CMB B-polarization measurements is that when the present ground-based and balloon-borne experiments are performed the sensitivity in the $\Omega_{gw}$-$f$ plot will have a one-order improvement to $10^{-16}$: when the proposed space missions are flown and completed the sensitivity will have another order of magnitude improvement to $10^{-17}$.

The instrument sensitivity goals of DECIGO[72], Big Bang Observer[73] and 6-S/C ASTROD-GW[68] all reach the $10^{-17}$-level in terms of $\Omega_{gw}$ (Fig. 4). The sensitivities of IPTA, FAST and SKA also reach the $10^{-17}$-level or beyond in terms of $\Omega_{gw}$ (Fig. 4). These instrument sensitivities are good enough to probe primordial GWs down to the $10^{-17}$-level or beyond in terms of $\Omega_{gw}$ at frequencies around 1 nHz, 10-300 μHz and 0.1-1 Hz to search and test inflationary/non-inflationary physics. The main issue is the level of foreground and whether foreground could be separated.

### 4.6. Very high frequency and ultrahigh frequency GW sources

There are four kind of potential GW sources in the very high frequency and ultrahigh frequency bands:[39]

(i) Discrete sources;[173]
(ii) Cosmological sources;[174]
(iii) Brane-world Kaluza–Klein (KK) mode radiation;[175,176]
(iv) Plasma instabilities.[177]

In general, objects do not radiate efficiently at wavelengths very different from their size.

This implies objects radiate at these bands need to be very small and yet have a very large energy concentration to induce significantly large curvature fluctuations. Grischuk[174] estimated the GWs generated from the amplification of quantum fluctuations by inflation. GWs in these bands with current wavelengths would had very short wavelengths that new physics might be working in the period of generation. However, the nucleosythesis bound of $\Omega_{gw}(f) \approx 10^{-5}$ must be satisfied by the spectrum of any GW background.[24] $h_c$ would at 100 MHz, 10 GHz and 1 THz would need to be less than $9.5 \times 10^{-29}$, $9.5 \times 10^{-31}$ and $9.5 \times 10^{-33}$, respectively. The actual signals may be much lower. Various theoretical models[178-185] predict GWs at levels from $\Omega_{gw}(f) \sim 10^{-8}$ to below $\sim 10^{-18}$. See [39] and references therein for more details.

To close this subsection, we quote from [39]: "Even assuming the most optimistic noise temperatures and the highest magnet strengths, detection of the cosmological signals look beyond reasonable extrapolation of current performance whereas very high-frequency GWs from brane-world scenarios may be within range of current technology. The most optimistic plasma instability signals from our galaxy if they occur at the low-frequency end of the range could also be above the sensitivity of future microwave detectors. ... There may also be astrophysical processes that convert violent electromagnetic events into very high-frequency gravitational sources that could be detected but more targeted modelling is needed to identify candidate astronomical objects. The technology for detectors which convert the GW directly to an electromagnetic signal is currently available and builds on decades of development for other applications."

### 4.7. Other possible sources

Cosmic strings are popular GW sources in many theoretical investigations. For possible GW magnitudes in various band of cosmic-string contribution, please see [186] and references therein. Recently, Geng, Huang and Lu[187] proposed the coalescence of strange-quark planets with strange stars as a new kind of GW burst sources for ground-based interferometers. As GW astronomy and GW physics progress, there could be detected GW sources of various different origins. This is open until the experiments and observations are performed. Possible GW sources which we have not discussed are GWs from cosmic strings, thermal GW radiation etc.

### 5. Discussion and Outlook

In spite of tremendous efforts in the high frequency band and some efforts in the very high frequency band experiments, gravitational wave has not been directly detected yet. This is due to the weakness in the strength of gravitational waves in the present epoch.

The first generation of km-sized arm length interferometers reach the sensitivity of detecting binary neutron star inspirals up to the Virgo cluster distance. From the statistics of astrophysical binary neutron star distribution, the rate of detection is about 0.05 events per year with a large uncertainty. However with a tenfold increase of strain sensitivity, the reach in distance increases by tenfold and the reach in astrophysical volume increases by one thousand fold. Hence the rate of detection is about 50 events per year. This is the goal of Advanced LIGO,[55] Advanced Virgo[56] and KAGRA/LCGT[57] under construction. Advanced LIGO has achieved 3 times better sensitivities with a reach to neutron star binary merging event at 70 Mpc and began its first observing run (O1) on September 18, 2015 searching for GWs. We could expect detection of GWs anytime. We will see a global network of second generation km-size interferometers for GW detection soon.

Another avenue for real-time direct detection is from the PTAs. The PTA bound on stochastic GW background already excludes most theoretical models; this may mean we could detect very low frequency GWs anytime too with a longer time scale.

We have presented a complete frequency classification of GWs according to their detection methods. Although there is no direct real-time detection of GWs yet, several bands are amenable to direct detection. Real-time direct detection may first come in the high frequency band or in the very low frequency band. Although the prospect of a launch of space GW is only expected in about 20 years, the detection in the low frequency band may have the largest signal to noise ratios. This will enable the detailed study of black hole co-evolution with galaxies and with the dark energy issue. Foreground separation and correlation detection method need to be investigated to achieve the sensitivities $10^{-16}$-$10^{-17}$ or beyond in $\Omega_{gw}$ to study the primordial GW background for exploring very early universe and possibly quantum gravity regimes.

When we look back at the theoretical and experimental development of GW physics and astronomy over the last 100 years, there are many challenges, some pitfalls, and during last 50 years close interactions among theorists and experimentalists. The subject and community have become clearly multidisciplinary. One example is the interaction of the GW community and the Quantum Optics community in the last 40 years to identify standard quantum uncertainties in measurement, to realize that this is not an obstacle of measurement in principle, and to find ways to overcome it. Another example is the interaction of the physics community and the astronomy community to understand and to identify detectable and potentially detectable GW sources. With current technology development and astrophysical understanding, we are in a position using GWs to study more thoroughly galaxies, supermassive black holes and clusters together with cosmology, and to explore deeper into the origin of gravitation and our universe. Next 100 years will be the golden age of GW astronomy and GW physics. The current and coming generations are holding such promises.

### Acknowledgements

This review extends and updates the former review [16]. It includes significant amount of materials from Ref.'s [42, 68, 69]. We would like to thank M. Bucher, A. Di Virgilio, N. Kanda, L. Lentati, R. N. Manchester, D. H. Reitze and L. Wen for their help and comments on various stages of writing. This work was supported in part by the National Science Council (Grant No. NSC102-2112-M-007-019), and by the MEXT (JSPS Leading-edge Research Infrastructure Program, JSPS Grant-in-Aid for Specially Promoted Research 26000005, MEXT Grant-in-Aid for Scientific Research on Innovative Areas 24103005, JSPS Core-to-Core Program, A. Advanced Research Networks, and the joint research program of the Institute for Cosmic Ray Research, University of Tokyo).